\documentstyle[11pt]{article}

\def\be{\begin{equation}}
\def\ee{\end{equation}}
\def\ba{\begin{eqnarray}}
\def\ea{\end{eqnarray}}
\def\bann{\begin{eqnarray*}}
\def\eann{\end{eqnarray*}}
\def\mref#1{Eq.~(\ref{eq:#1})}
\def\nref#1{(\ref{eq:#1})}
\def\oref#1{\ref{eq:#1}}

\def\mlab#1{\label{eq:#1}}
\def\nn{\nonumber}
\def\Comment#1{}

\def\im{i}

\def\l{\left}
\def\r{\right}
\def\logten{\log_{10}}

\def\norm#1{\l\|#1\r\|}

\def\GeV{\mbox{GeV}}

\def\half{\mbox{\small{$\frac{1}{2}$}}}

\def\mfrac#1#2{\mbox{\small{$\frac{#1}{#2}$}}}

\begin{document}

\baselineskip=20pt

\hfill RUGTh-971219

\begin{center}
{\LARGE \bf
Running Coupling in Non-Perturbative QCD\\
}
\vspace{6mm}
{\Large I. Bare Vertices and y-max Approximation
}
\vspace{6mm}

{\bf D. Atkinson\footnote{atkinson@phys.rug.nl} and
J.C.R. Bloch\footnote{bloch@phys.rug.nl}\\
Institute for Theoretical Physics\\
University of Groningen\\
Nijenborgh~4\\
NL-9747~AG~~Groningen\\
The Netherlands
}
\end{center}

{\bf Abstract}
\begin{center}
{\parbox{120mm}{\small A recent claim that in quantum chromodynamics
the gluon propagator vanishes in the infrared limit, while the ghost
propagator is more singular than a simple pole, is investigated analytically
and numerically.  This picture is shown to be supported even at the
level in which the vertices in the Dyson-Schwinger equations are taken
to be bare. The running coupling is shown to be uniquely determined by the
equations and to have a large finite infrared limit.   }}
\end{center}

\section{Introduction}

The proof of the renormalizability of non-Abelian gauge theories like
QCD\cite{GtH}, and the discovery of ultraviolet asymptotic 
freedom\cite{Politzer}, heralded a new phase in the acceptance of quantum
field theories as serious candidates for the quantitative description
of the weak, electromagnetic and strong interactions. Since the
running coupling in QCD decreases logarithmically to zero as the
renormalization point is taken to infinity, it seems reasonable to
calculate it perturbatively in the deep ultraviolet regime, where it is
very small, even though a proof is lacking that the perturbation
series makes sense (for example, that it is strongly asymptotic). 

Although one is not sure that perturbation theory is reliable for QCD
at very high energies, at very low energies it is quite clear that it
is inadequate. Chiral symmetry breaking and fermion mass generation
are typically non-perturbative phenomena. The obverse of ultraviolet
asymptotic freedom is infrared slavery or confinement. Since the
coupling decreases as the energy increases, it increases as one goes
to lower energies, and the possibility is open that its infrared
limit is infinite. Many attempts\cite{Roberts}
 --- necessarily of a non-perturbative
nature --- have been made to show this divergence of the coupling in
the infrared limit. Mandelstam initiated the study of the gluon
Dyson-Schwinger equation in Landau gauge\cite{Mandelstam}. Although he
did consider the gluon-ghost coupling, Mandelstam concluded provisionally that
its effect could safely be neglected. This assumption was also made in
subsequent work\cite{DA, MRP}. A deficiency of these
attempts to show that the gluon propagator is highly singular in the
infrared is the necessity to posit certain 
cancellations of leading terms in the equations. An uncharitable case
of {\em petitio principii} might almost be made (i.e. circularity).

Recently, a new possibility has been opened up by the work of von
Smekal, Hauck and Alkofer\cite{Smekal}. In this work the coupling of
the gluon to the ghost was not
neglected. These authors claim
that it is not the gluon, but rather the ghost propagator that is
highly singular in the infrared limit. The running coupling itself
has a {\em finite} though quite large value in the limit of zero
energy, presumably large enough to guarantee chiral symmetry breaking in
the quark equation\cite{chiral}. 

In the present paper we investigate the claims made in the new work. 
We shall write the gluon propagator in Landau gauge as 
\[
D^{ab}_{\mu\nu}(p)=
-\delta^{ab}\frac{1}{p^2}
\Delta_{\mu\nu}(p)  F(-p^2)\, ,
\]
where $a$ and $b$ are colour indices, 
and where $\Delta =\Delta^2$ is the projection operator 
\[
\Delta_{\mu\nu}(p)  = g_{\mu\nu} - \frac{p_\mu p_\nu}{p^2} \, .
\]
The ghost propagator will be written in the form  
\[
G^{ab}(p)=-\delta^{ab}\frac{1}{p^2}G(-p^2)\, ,
\]
and we shall refer to the scalar functions $F$ and $G$ as the gluon
and ghost form factors, respectively. 

The claim made in Ref.\cite{Smekal} is that, in the infrared limit 
$x=-p^2\rightarrow 0$, these form factors have the following
behaviour:
\be
\mlab{smekal2}
F(x)\sim x^{2\kappa}\hspace{1cm}G(x)\sim x^{-\kappa}\, ,
\ee
where $\kappa \approx 0.92$. To obtain these results certain Ans\"atze
were made for the three-gluon and ghost-gluon vertices, functional
forms inspired, but not uniquely determined by Slavnov-Taylor
identities. In fact the Ansatz made for the ghost-gluon vertex is such
that actually the infrared behaviour \mref{smekal2} is not consistent
with the Dyson-Schwinger equations. The difficulty is the occurrence
of a term 
\be
\mlab{logp}
\int^{\Lambda^2}_x\frac{dy}{y}F(y)G^2(y)
\ee
in the equation for the ghost form factor, which, with the form
\nref{smekal2}, would yield an impermissible $\log x$ factor in the
limit $x\rightarrow 0$. Von Smekal et al. circumvent this problem by
replacing {\em one} of the factors $G(y)$ by $G(x)$, thereby undermining to
a large extent their supposed improvement of the vertex Ansatz. 

Since we found the {\em ad hoc} nature of this last replacement
questionable, we decided first to see what would happen if one simply
replaces the full vertices by bare ones. In this case the problematic
logarithm of \mref{logp} does not occur, and we can simply analyze the
equation as it stands. If the behaviour \nref{smekal2} were to go
away, it would bode ill for the new approach. However, our finding is
that, with bare vertices, the form \nref{smekal2} indeed remains good,
but with the index changed to $\kappa \approx 0.77$. Moreover, we can
show that the solutions of the coupled gluon and ghost equations lie
on a three-dimensional manifold, i.e. the general solution has three
free parameters; nevertheless all solutions have the infrared
behaviour \nref{smekal2}. Our primary purpose in this initial paper is
to explain the above findings in detail.

In Landau gauge, the QCD Dyson-Schwinger 
equations lead to the following coupled integral 
equations  for the renormalized 
gluon and ghost form factors:
\ba
F^{-1}(p^2)=Z_3 &+& \frac{g^2}{8\pi^3} \tilde{Z}_1
\int_0^{\Lambda^2}\frac{dq^2}{p^2}G(q^2)\int_0^\pi d\theta 
\sin^2\theta \, M(p^2,q^2,r^2)G(r^2) \nn\\ 
&+&\frac{g^2}{8\pi^3} Z_1 
\int_0^{\Lambda^2}\frac{dq^2}{p^2}F(q^2)\int_0^\pi d\theta 
\sin^2\theta \, Q(p^2,q^2,r^2)F(r^2) \nn\\
\mlab{ghostda}
G^{-1}(p^2) = \tilde{Z_3}
&-& \frac{3g^2}{8\pi^3} \tilde{Z}_1
\int_0^{\Lambda^2}dq^2q^2G(q^2)\int_0^\pi d\theta 
\frac{\sin^4\theta}{r^4}F(r^2) \, ,
\ea
with $r^2=p^2+q^2-2pq\cos\theta$. The kernels are  
\[
M(p^2,q^2,r^2) = 
\frac{1}{r^2}\left(
\frac{p^2+q^2}{2}-\frac{q^4}{p^2}\right) +\frac{1}{2}+
\frac{2q^2}{p^2}-\frac{r^2}{p^2}
\, .
\]
\bann Q(p^2,q^2,r^2) &=&
\l(\frac{p^6}{4q^2} + 2p^4 - \frac{15q^2p^2}{4} + \frac{q^4}{2} 
+ \frac{q^6}{p^2}\r)\frac{1}{r^4}
+\l(\frac{2p^4}{q^2} - \frac{19p^2}{2} - \frac{13q^2}{2} 
+ \frac{8q^4}{p^2}\r)\frac{1}{r^2}\\
&& -\l(\frac{15p^2}{4q^2} + \frac{13}{2} + \frac{18q^2}{p^2}\r)
+\l(\frac{1}{2q^2} + \frac{8}{p^2}\r)r^2 + \frac{r^4}{p^2 q^2} \, .
\eann
Here the full three-gluon and the ghost-gluon vertices
have been replaced by their bare values, while the four-gluon and
quark-gluon vertices have been provisionally thrown away.  To obtain
these equations from the Dyson-Schwinger equations, we performed a
Wick rotation and evaluated two trivial angular integrations. 

The form factors and the QCD coupling are renormalized using some
renormalization prescription, $Z_1$, $Z_3$, $\tilde Z_1$, $\tilde Z_3$ 
being the renormalization constants for the triple gluon vertex, the gluon
field, the gluon-ghost vertex and the ghost field defined by
\be
\overline F(p^2) = Z_3 F(p^2)\hspace{1cm}
\overline G(p^2) = \tilde Z_3 G(p^2)\hspace{1cm}
g = \frac{Z_3^{3/2}}{Z_1} g_0 = \frac{\sqrt{Z_3}\tilde
Z_3}{\tilde{Z}_1} g_0 \,,
\mlab{Zs}
\ee
where $\overline F(p^2)$, $\overline G(p^2)$ are the unrenormalized
gluon and ghost form factors, $F(p^2)$, $G(p^2)$ the renormalized
ones, $g_0$ is the bare coupling and $g$ its renormalized value.
One usually writes the renormalization constants and the renormalized
quantities as functions of a renormalization scale $\mu$,
corresponding to a specific renormalization prescription. 
However, we will see in the following sections that the
renormalization prescriptions can be made more general than is usually
done in perturbation theory, and that each prescription will
correspond to a solution of the non-perturbative integral
equations. Instead of having the usual invariance of the running
coupling with respect to a variation of the
renormalization scale, we will find a more general invariance under an
arbitrary transformation in the three-dimensional space of solutions
of the integral equations.

We wish to solve the coupled integral equations \nref{ghostda} for $F$
and $G$, and we propose to do that in a future publication. For the
moment we introduce a further simplification, the y-max
approximation. This amounts to replacing $F(r^2)$ and $G(r^2)$ in
\mref{ghostda} by $F(p^2)$ and $G(p^2)$ if $p^2>q^2$, but by  
$F(q^2)$ and $G(q^2)$ if $p^2\le q^2$. This approximation facilitates the
analytical and numerical analysis of the equations, since the angular
integrals can now be performed exactly, and indeed the resulting
one-dimensional Volterra equations can be converted into nonlinear
ordinary differential equations. This y-max approximation is very
widely employed for these reasons, in particular by von Smekal et al.,
and although we do the same thing here, let us sound a note of
warning: although we do not expect the qualitative picture of
\mref{smekal2} to change, we do expect the value of the index $\kappa$
to be different when we treat the coupled equations without the y-max
approximation. We have already seen that $\kappa$ is sensitive to the
choice of Ansatz for the vertex functions, and it is also affected by
the y-max approximation. The bare vertex Ansatz is of course only a
first guess; and it is clear also that the Ansatz of von Smekal et
al. needs to be improved (the logarithm problem to which we alluded
above persists when one no longer employs the y-max approximation).
Nevertheless, the picture that von Smekal, Hauck and Alkofer have
uncovered appears to be robust in its qualitative, and hopefully also
in its semi-quantitative features: the gluon propagator is {\em soft}
in the infrared (i.e. it vanishes in this limit, instead of blowing
up like a pole), while the ghost propagator is {\em hard} (it is more
singular than a pole). The consequences for the physics of the strong
interaction need to be investigated.

\section{The coupled gluon-ghost equations}

The set of coupled integral equations for the gluon and ghost propagator,
using the bare triple gluon vertex and the bare gluon-ghost vertex, and
introducing the y-max approximation, is as follows:
\ba
F^{-1}(x) &=& Z_3 
+ \lambda \tilde Z_1 \l[ 
G(x)\int_0^x \, \frac{dy}{x} \, \l(-\frac{y^2}{x^2}+\frac{3y}{2x}\r) \, G(y)
+ \int_x^{\Lambda^2} \, \frac{dy}{2y} \, G^2(y) \r]
\mlab{Feq} \\
&& \hspace{6mm} + \lambda Z_1 \l[ 
F(x)\int_0^x \, \frac{dy}{x} \,
\l(\frac{7y^2}{2x^2}-\frac{17y}{2x}-\frac{9}{8}\r) \, F(y) 
+ \int_x^{\Lambda^2} \, \frac{dy}{y} \, \l( -7 + \frac{7x}{8y}\r) \, F^2(y)
\r] \nn\\[3mm]
G^{-1}(x) &=& \tilde Z_3
- \frac{9}{4} \lambda \tilde Z_1 \l[
F(x)\int_0^x \, \frac{dy}{x} \, \frac{y}{x} \, G(y)
+ \int_x^{\Lambda^2} \frac{dy}{y} \, F(y)G(y) \r] \,,
\mlab{Geq}
\ea
where $\lambda=g^2/16\pi^2$, $x=p^2$ and $y=q^2$.

To solve Eqs.~(\oref{Feq}, \oref{Geq}), we eliminate the renormalization
constants $Z_3$ and $\tilde Z_3$ by subtracting the equations at 
$x=\sigma$:
\ba
F^{-1}(x) &=& F^{-1}(\sigma) \mlab{Feqsub} \\
&& \hspace{-1.8cm} + \lambda Z_1 \l[ 
F(x)\int_0^x \, \frac{dy}{x} \,
\l(\frac{7y^2}{2x^2}-\frac{17y}{2x}-\frac{9}{8}\r) \, F(y)
- F(\sigma)\int_0^\sigma \, \frac{dy}{\sigma} \,
\l(\frac{7y^2}{2\sigma^2}-\frac{17y}{2\sigma}-\frac{9}{8}\r) \, F(y) \r. \nn \\
&& \l. - 7 \int_x^\sigma \, \frac{dy}{y} \,  \, F^2(y)
+ \int_x^{\Lambda^2} \, \frac{dy}{y} \, \l( \frac{7x}{8y}\r) \, F^2(y)
- \int_\sigma^{\Lambda^2} \, \frac{dy}{y} \, \l( \frac{7\sigma}{8y}\r) \, 
F^2(y) \r] \nn\\
&& \hspace{-1.8cm} + \lambda \tilde Z_1 \l[ 
G(x)\int_0^x \, \frac{dy}{x} \, \l(-\frac{y^2}{x^2}+\frac{3y}{2x}\r) \, G(y)
- G(\sigma)\int_0^\sigma \, \frac{dy}{\sigma} \,
\l(-\frac{y^2}{\sigma^2}+\frac{3y}{2\sigma}\r) \, G(y) \r. \nn \\
&& \l. + \int_x^{\sigma} \, \frac{dy}{2y} \, G^2(y) \r] \nn \\[3mm]
\mlab{Geqsub}
G^{-1}(x) &=& G^{-1}(\sigma) \\
&& - \frac{9}{4} \lambda \tilde Z_1 \l[
F(x)\int_0^x \, \frac{dy}{x} \, \frac{y}{x} \, G(y) 
- F(\sigma)\int_0^\sigma \, \frac{dy}{\sigma} \, \frac{y}{\sigma} \, G(y)
 + \int_x^\sigma \frac{dy}{y} \, F(y)G(y) \r] \,. \nn
\ea

\section{Symmetries of the reduced equations}
\label{Sect:symmetries}

A very interesting simplification of Eqs.~(\oref{Feqsub}, \oref{Geqsub}) is
obtained if we throw away the gluon loop in \mref{Feqsub}, keeping only the
ghost loop.  This truncation is particularly interesting because, as we
will show, its properties agree qualitatively with the requirements of a
consistent physical picture, whereas the inclusion of the gluon loop
introduces an ambiguity, which is probably due to the presence of terms
involving the yet unknown renormalization constant $Z_1$. The
truncated set of equations is:
\ba
\mlab{Feqsub3} 
F^{-1}(x) &=& F^{-1}(\sigma) \\ 
&& \hspace{-1.8cm} + \lambda \tilde Z_1 \l[
G(x)\int_0^x \, \frac{dy}{x} \, \l(-\frac{y^2}{x^2}+\frac{3y}{2x}\r) \,
G(y) - G(\sigma)\int_0^\sigma \,
\frac{dy}{\sigma} \, \l(-\frac{y^2}{\sigma^2}+\frac{3y}{2\sigma}\r) \, G(y)
+ \int_x^{\sigma} \, \frac{dy}{2y} \, G^2(y) \r] \nn \\[5mm]
G^{-1}(x)
&=& G^{-1}(\sigma) - \frac{9}{4} \lambda \tilde Z_1 \l[
F(x)\int_0^x \, \frac{dy}{x} \, \frac{y}{x} \, G(y) -
F(\sigma)\int_0^\sigma \, \frac{dy}{\sigma} \, \frac{y}{\sigma} \, G(y) +
\int_x^\sigma \frac{dy}{y} \, F(y)G(y) \r] \,. \nn\\
\mlab{Geqsub3}
\ea

We will show that Eqs.~(\oref{Feqsub3}, \oref{Geqsub3}) have a
three-dimensional space of solutions and that these solutions can be
transformed into one another by means of simple scalings.

First of all, if we have a solution $F(x)$ and $G(x)$, we can build a
two-dimensional infinity of solutions simply by scaling these
functions:
\ba
\tilde{F}(x) &=& F(x)/a \mlab{scala} \\
\tilde{G}(x) &=& G(x)/b \mlab{scalb}
\ea
which simply amounts to a redefinition of $Z_3$ and $\tilde{Z}_3$,
i.e. to a change in the renormalization prescription. The new
functions satisfy the same integral equations, with the rescaled
coupling constant:
\[
\tilde{\lambda}=\lambda ab^2\, .
\]
Although the value of $\lambda$ is in general changed, this has no
physical significance, since the following gauge invariant quantity is
unchanged by the above transformations:
\[
\lambda F(x)G^2(x)=\tilde{\lambda} \tilde{F}(x)\tilde{G}^2(x)\, .
\]
Thus the two-dimensional manifold of solutions corresponds to the same
physics, and one could, for instance, take $\lambda =1$ and set 
$F(\sigma )=1$ without essential loss of generality.

A second, less trivial feature is the possibility to derive an
infinite number of solutions starting from $F(x)$ and $G(x)$ just by
scaling the momentum $x$ to $tx$. The new functions $\hat F$ and $\hat
G$ take the same values at momentum $x$ as $F$ and $G$ at momentum
$tx$: 
\be 
\hat F(x) \equiv F(tx)
\hspace{1.5cm}
\hat G(x) \equiv G(tx) \, .
\mlab{scalt}
\ee

In terms of the scaled quantities,
\[
\tilde x = x/t
\hspace{1cm}
\tilde y = y/t
\hspace{1cm}
\tilde\sigma = \sigma/t
\]
we find
\bann
\hat F^{-1}(\tilde{x}) &=& \hat F^{-1}(\tilde\sigma) \\
&& \hspace{-1.8cm} + \lambda \tilde Z_1 \l[ 
\hat G(\tilde{x})\int_0^{\tilde{x}} \, \frac{d\tilde{y}}{\tilde{x}} \,
\l(-\frac{\tilde{y}^2}{\tilde{x}^2}+\frac{3\tilde{y}}{2\tilde{x}}\r)
\, \hat G(\tilde{y}) - \hat G(\tilde\sigma)\int_0^{\tilde\sigma} \, 
\frac{d\tilde{y}}{\tilde\sigma} \,
\l(-\frac{\tilde{y}^2}{\tilde\sigma^2}+\frac{3\tilde{y}}{2\tilde\sigma}\r)
\, \hat G(\tilde{y}) + \int_{\tilde{x}}^{\tilde\sigma} \,
\frac{d\tilde{y}}{2\tilde{y}} \, \hat G^2(\tilde{y}) \r] \\[5mm]
\hat{G}^{-1}(\tilde{x})
&=& \hat{G}^{-1}(\tilde{\sigma}) - \frac{9}{4} \lambda \tilde Z_1 \l[
\hat{F}(x)\int_0^{\tilde{x}} \, \frac{d\tilde{y}}{\tilde{x}} \,
\frac{\tilde{y}}{\tilde{x}} \, \hat{G}(\tilde{y}) -
\hat{F}(\tilde{\sigma})\int_0^{\tilde{\sigma}} \,
\frac{d\tilde{y}}{\tilde{\sigma}} \, \frac{\tilde{y}}{\tilde{\sigma}}
\, \hat{G}(\tilde{y}) + \int_{\tilde{x}}^{\tilde{\sigma}}
\frac{d\tilde{y}}{\tilde{y}} \, \hat{F}(\tilde{y})\hat{G}(\tilde{y}) \r] \,.
\eann

This means that $\hat F(x)$ and $\hat G(x)$ are also solutions of the
integral equations solved by $F(x)$ and $G(x)$.  Again, all the
solutions obtained by varying the scaling factor $t$ correspond to the
same physical picture, since a scaling of momentum merely corresponds to
choosing the units for the momentum variable when renormalizing the
coupling constant at a certain physical scale. It is clear that the
three above-mentioned scaling properties allow us to construct the
whole three-dimensional space of solutions starting from one specific
solution.

This three-fold scaling invariance has a physical relevance, as we will now
show. From the renormalization of the gluon-ghost-ghost vertex, we
define the renormalized coupling as
\be
\alpha = \tilde Z_1^{-2} Z_3 \tilde Z_3^2 \, \alpha_0 \,,
\mlab{gGG}
\ee

where $\alpha_0=g_0^2/4\pi$. According to Taylor\cite{Taylor}, 
$\tilde Z_1=1$ in the Landau
gauge and, writing \mref{gGG} with two different renormalization
prescriptions, we find
\be
\alpha Z_3^{-1} \tilde Z_3^{-2} = \hat\alpha \hat{Z}_3^{-1}
\hat{\tilde{Z}}_3^{-2} \,.
\ee

Substituting \mref{Zs} for $Z_3$ and $\tilde Z_3$ and eliminating the
unrenormalized quantities, we have
\be
\alpha F(x) G^2(x) = \hat\alpha \hat{F}(x) \hat{G}^2(x) \,.
\ee

This allows us to define the running coupling
\be
\alpha(x) = \alpha F(x) G^2(x) \,,
\mlab{rc}
\ee
which is independent of the renormalization prescription.

This invariance of $\alpha(x)$ with respect to the renormalization
prescription is exactly reproduced 
in the {ghost-loop-only} truncation. Furthermore, the renormalization
prescription is more general than what is normally used in
perturbation theory. There the prescription usually is
$F(\mu)=G(\mu)=1$ and $\alpha=\alpha_\mu^{\mbox{exp}}$, and the
invariance is taken to be an invariance with respect to the choice of
the point 
$\mu$ at which these conditions are imposed. In our non-perturbative
treatment the invariance is with respect to an arbitrary
transformation in the three-dimensional space of solutions of the
equations. For many of these solutions, there is even no such no such
point at which 
$F(\mu)=G(\mu)=1$ or where $\alpha=\alpha_\mu^{\mbox{exp}}$, so
that no traditional scale $\mu$ can be attached to the renormalization
prescription itself: only the running coupling $\alpha(x)$ defined in
\mref{rc} is physically meaningful. We will see later that the loss of
symmetry of the equations when we include the gluon loop in the
current truncation scheme destroys this invariance with respect to an
arbitrary renormalization prescription, and different prescriptions 
no longer lead to the same physical running coupling.

\section{Infrared behaviour}
\label{Sect:IR}

We will show analytically that the equations Eqs.~(\oref{Feqsub},
\oref{Geqsub}) and Eqs.~(\oref{Feqsub3}, \oref{Geqsub3}) have a consistent
infrared asymptotic solution:
\ba
F(x) &=& A x^{2\kappa} \mlab{FIR}\\
G(x) &=& B x^{-\kappa} \mlab{GIR} \, ,
\ea
and that these solutions even solve the {\it ghost-loop-only}
equations~(\oref{Feqsub3}, \oref{Geqsub3}) exactly for all momenta.

Let us try the Ansatz
\be
\mlab{sol}
F(x)=Ax^\alpha \hspace{1cm} G(x)=Bx^\beta\, . \mlab{FGIR}
\ee
In the infrared asymptotic regime the gluon loop does not contribute to
lowest order. Substituting \mref{FGIR} into the integral
equations~(\oref{Feqsub3}, \oref{Geqsub3}) we calculate
\be
A^{-1}x^{-\alpha}=A^{-1}\sigma^{-\alpha}+\lambda \tilde Z_1 B^2
\left[
\frac{3}{2}\frac{1}{2+\beta}-\frac{1}{3+\beta}-\frac{1}{4\beta}
\right]
   (x^{2\beta}-\sigma^{2\beta})
\mlab{aaa}
\ee
and
\be
B^{-1}x^{-\beta}=B^{-1}\sigma^{-\beta}-\frac{9}{4}\lambda \tilde Z_1 A B
\left[
\frac{1}{2+\beta}+\frac{1}{\alpha +\beta}
\right] (x^{\alpha +\beta}-\sigma^{\alpha +\beta})
\mlab{bbb}
\ee
on condition that
\be
\beta >-2  \mlab{betamin}
\ee
to avoid infrared singularities. The powers on both sides of
Eqs.~(\oref{aaa}, \oref{bbb}) agree if
\[
\alpha =-2\beta \, ,
\]
and defining the index $\kappa$ by
\be
\alpha = 2\kappa \hspace{1cm} \beta=-\kappa 
\ee 
we find that both the constant and the power terms in \mref{aaa} and
\mref{bbb} match if
\ba
\lambda \tilde Z_1 A B^2 &=& \l[\frac{3}{2(2-\kappa)} - \frac{1}{3-\kappa}
+ \frac{1}{4\kappa}\r]^{-1} \mlab{AB2-1}\\[3mm]
\hspace{-1cm}\mbox{and}\hspace{1cm}
\lambda \tilde Z_1 A B^2 &=& - \frac{4}{9} \l[\frac{1}{2-\kappa}
- \frac{1}{\kappa} \r]^{-1} \, .
\mlab{AB2-2}
\ea
Elimination of $\lambda \tilde Z_1 A B^2$ yields a quadratic equation
for $\kappa$, which remarkably does not depend on the value of the coupling
strength $\lambda$:
\be
19\kappa^2 + 77\kappa + 48 = 0 \,,
\ee
which has two real solutions
\be
\mlab{kappa2}
\kappa = \frac{77 \pm \sqrt{2281}}{38},
\ee
or
\be
\kappa_1 \approx 0.769479 \hspace{1cm}\mbox{and}\hspace{1cm} 
\kappa_2 \approx 3.28315.
\ee

The second root is spurious: it must be rejected because it gives rise to
infrared singularities and thus does not give a solution of the integral
equation.

Replacement of $\kappa$ by $\kappa_1$ in \mref{AB2-1} or \mref{AB2-2}
yields the condition:
\be
\nu = \lambda \tilde Z_1 A B^2 \approx 0.912771 \,.
\mlab{nu}
\ee

From \mref{rc} we know that the running coupling is given by
\be
\alpha(x) = 4 \pi \lambda F(x) G^2(x)
\ee
in the Landau gauge.
Condition \mref{nu} is important, as it tells us that the running coupling
has a non-trivial infrared fixed point
\be
\lim_{x\to 0} \alpha(x) \approx 11.4702 \, .
\ee
This means that the ghost field, which only introduces quantitative
corrections to the perturbative ultraviolet behaviour of the running
coupling, does alter its infrared behaviour in a very drastic way.

We will show further on that the running coupling remains almost constant
up to a certain momentum scale $\tilde x$, after which it decreases as
$1/\log{x}$. The momentum scale at which the constant bends over into a
logarithmic tail is closely related to the value of $\Lambda_{QCD}$. This
is easily understood intuitively, since the perturbative ultraviolet
behaviour of the running coupling blows up very quickly as the
momentum gets down to ${\cal O}(\Lambda_{QCD})$.

\section{Infrared asymptotic solution}
\label{Sect:irasexp}

Although we have seen in the previous section that the pure power
behaviours for $F(x)$ and $G(x)$ solve the reduced equations exactly,
these power solutions only give rise to a two-dimensional space
of solutions.  However, the numerical results told us that the
equations were much richer then we initially believed. These numerical
results tended to suggest that the power solutions are only one very
special two-dimensional family of solutions in the midst of a whole
three-dimensional space. Typical non-power solutions showed an infrared
behaviour completely consistent with the power solution mentioned
earlier, which then bends over quite rapidly at some momentum $\tilde
x$ into a completely different ultraviolet behaviour which seemed to
be proportional to some power of the logarithm of momentum. A
straightforward investigation of the ultraviolet asymptotic behaviour
of the solutions tells us that such powers of logarithms are indeed
consistent ultraviolet solutions, but no obvious mechanism seemed
available to match the infrared to the ultraviolet parts of the
solutions, making us believe at first that the numerical program was
giving us spurious pseudo-solutions, due to some numerical
inaccuracies or artifacts. One of the main reasons was that the
infrared power behaviour only contains one free parameter, and a
standard asymptotic expansion does not add any corrections to the
leading power. If the infrared asymptotic solution contains only one
parameter, it was very unclear how an infinite number of solutions
with log-tails could develop out of each power solution. Nevertheless
the numerical results indicated that each power solution had an
infinite number of corresponding log-tailed solutions, and each
solution seemed to be characterized by the momentum at which the
log-tail sets in.

The traditional asymptotic expansion one would normally try, is as follows:
\ba
F(x) &=& x^{2\kappa} \sum_{i=0}^{N} A_i x^i \\
G(x) &=& x^{-\kappa} \sum_{i=0}^{N} B_i x^i .
\ea

The reason for this is that each term in the expansion usually generates
terms, through integration, that are of the same power or one unit
higher. However, the fact that the equations under consideration are {\it
exactly} solved by the power solution alters the reasoning. The leading
power term does not generate additional, next-to-leading order terms, and
all $A_i, B_i$ for $i>0$ have to be zero for consistency reasons.

However, the fact that the power solution solves the integral equations
does not mean that this is the unique solution, and we next tried an
infrared asymptotic solution of the shape:
\ba
F(x) &=& A_0 x^{2\kappa} + A_1 x^{\alpha_1} \\
G(x) &=& B_0 x^{-\kappa} + B_1 x^{\beta_1} \nn.
\ea
with $\alpha_1 > 2\kappa$ and $\beta_1> -\kappa$. Substitution of these
solutions into Eqs.~(\oref{Feqsub3}, \oref{Geqsub3}), tells us that
consistency is obtained if $\alpha_1 - \beta_1 = \kappa$, as for the
leading power, but it gives an additional constraint, fixing the value of
the exponent of the next-to-leading exponent. However, the solution proposed
above does generate additional higher order terms, and consistent
asymptotic infrared expansions can be built as follows:
\ba
F(x) &=& x^{2\kappa} \sum_{i=0}^{N} A_i x^{i\rho} \mlab{asexp} \\
G(x) &=& x^{-\kappa} \sum_{i=0}^{N} B_i x^{i\rho} \nn \,,
\ea
where the exponents of successive powers always increase by the same amount
$\rho > 0$. To check the consistency of these infrared asymptotic
expansions, we substitute them into Eqs.~(\oref{Feqsub3},
\oref{Geqsub3}). We make a Taylor expansion of the left-hand sides of these
equations and expand the series multiplications, before integration, on the
right-hand sides. Consistency requires that the coefficients of equal
powers of momentum match each other on both sides of the equations.

The conditions on the leading term remain unchanged as described in
Sect.~\ref{Sect:IR}, with $\kappa \approx 0.769479$ and $\nu=\lambda\tilde
Z_1 A_0 B_0^2 \approx 0.912771$. Equating the second order terms on left and
right-hand sides of both integral equations yields the following set of two
{\it homogeneous} linear algebraic equations for $a_1 \equiv A_1/A_0$ and
$b_1 \equiv B_1/B_0$:
\ba
&& \frac{a_1}{\nu}+\l(\frac{3}{2(2-\kappa+\rho)}-\frac{1}{3-\kappa+\rho}
+\frac{3}{2(2-\kappa)}-\frac{1}{3-\kappa}-\frac{1}{-2\kappa+\rho}\r) b_1 =
0 \nn\\
&& \l(\frac{1}{\kappa+\rho}-\frac{1}{2-\kappa}\r) a_1
+ \l(\frac{1}{\kappa+\rho}-\frac{1}{2-\kappa+\rho}+\frac{4}{9\nu}\r) b_1 =
0\nn
\ea

This set of equations will only have non-trivial solutions if its
determinant is zero, in which case it will have a one-parameter infinite
number of solutions. The characteristic equation is:
\be
-9.27685 \, \rho^4 - 15.5544 \, \rho^3 + 30.2899 \, \rho^2
+ 71.5686 \,\rho = 0 \, .
\ee

The four solutions are:
\be
\rho = 0 \,, \hspace{1cm}
\rho = 1.96964 \,, \hspace{1cm}
\rho = -1.82316 \pm 0.770012 \, \im \,.
\ee

The solution $\rho=0$ corresponds to the pure power solution. The two
complex solutions are spurious as they are not consistent with
$\mbox{Re}\,\rho>0$, while the solution $\rho = 1.96964$ gives rise to
consistent infrared asymptotic expansions.

The linear homogeneous set of equations then yields
\be
\eta \equiv b_1/a_1 = 0.829602 \, ,
\mlab{eta}
\ee
and the solutions of this set of equations can, for instance, be
parametrized by $a_1$.

Let us define
\be
a_n = A_n / A_0 \,, \hspace{1.5cm}
b_n = B_n / B_0 \, ,\nn
\ee
in terms of which we find the
following {\it heterogeneous} set of equations for $a_2$ and $b_2$:
\ba
\mlab{a2b2}
&&\frac{a_2}{\nu} +
\l[\frac{3}{2(2-\kappa)}-\frac{1}{3-\kappa}+\frac{3}{2(2-\kappa+2\rho)}
-\frac{1}{3-\kappa+2\rho}-\frac{1}{-2\kappa+2\rho} \r] b_2 \\
&& \hspace{4cm} = \frac{a_1^2}{\nu} -
        \l(\frac{3}{2(2-\kappa+\rho)}-\frac{1}{3-\kappa+\rho}
        -\frac{1}{2(-2\kappa+2\rho)}\r) b_1^2 \nn\\
&&\l[\frac{1}{\kappa+2\rho}-\frac{1}{2-\kappa}\r] a_2
+ \l[\frac{1}{\kappa+2\rho}-\frac{1}{2-\kappa+2\rho}+\frac{4}{9\nu}\r] b_2 
\nn\\
&& \hspace{4cm} = \frac{4 b_1^2}{9\nu}- \l(\frac{1}{\kappa+2\rho}
-\frac{1}{2-\kappa+\rho}\r) a_1 b_1 \, ,\nn
\ea
with unique solution
\[
a_2 = 0.408732 \; a_1^2 \hspace{1.5cm}
b_2 = 1.31169 \; a_1^2\nn
\]
and for $a_3, b_3$:
\ba
\mlab{a3b3}
&&\frac{a_3}{\nu} +
\l[\frac{3}{2(2-\kappa)}-\frac{1}{3-\kappa}+\frac{3}{2(2-\kappa+3\rho)}
-\frac{1}{3-\kappa+3\rho}-\frac{1}{-2\kappa+3\rho} \r] b_3 \\
&& \hspace{0.2cm} = \frac{2 a_1 a_2 - a_1^3}{\nu} -
        \l[\frac{3}{2(2-\kappa+2\rho)}-\frac{1}{3-\kappa+2\rho}
           \frac{3}{2(2-\kappa+\rho)}-\frac{1}{3-\kappa+\rho}
        -\frac{1}{-2\kappa+3\rho}\r] b_1 b_2 \nn\\
&&\l[\frac{1}{\kappa+3\rho}-\frac{1}{2-\kappa}\r] a_3
+ \l[\frac{1}{\kappa+3\rho}-\frac{1}{2-\kappa+3\rho}+\frac{4}{9\nu}\r] b_3 \nn\\
&& \hspace{1cm} = \frac{4 (2 b_1 b_2 - b_1^3)}{9\nu}
- \l(\frac{1}{\kappa+3\rho} -\frac{1}{2-\kappa+2\rho}\r) a_1 b_2 
- \l(\frac{1}{\kappa+3\rho} -\frac{1}{2-\kappa+\rho}\r) a_2 b_1 \, .\nn
\ea
with unique solution
\[
a_3 = -0.761655 \; a_1^3 \hspace{1.5cm}
b_3 = 0.783905 \; a_1^3\nn \, .
\]

By induction one can prove that the higher
order terms all yield sets of equations of the same nature as
\mref{a2b2} and \mref{a3b3}, where the right-hand side of the set defining
the coefficients $a_n, b_n$ are proportional to $a_1^n$. This means
that we have a general solution for the $n$th order coefficient of
the type
\be
a_n = f_n a_1^n \hspace{1.5cm}
b_n = g_n a_1^n 
\ee
for $n>1$, where the $f_n, g_n$ are constants (independent of
$\lambda$ and of $\tilde Z_1$).

The asymptotic expansions \mref{asexp} can thus be written in the form
\ba
F(x) &=& A_0 x^{2\kappa} \l(1 + \sum_{i=1}^{N} f_i a_1^i x^{i\rho}\r)
\mlab{asexp2} \\
G(x) &=& B_0 x^{-\kappa} \l(1 + \sum_{i=1}^{N} g_i a_1^i x^{i\rho}\r) \,, \nn
\ea
where $A_0$, $B_0$ and $a_1=A_1/A_0$ are chosen to be the free
parameters spanning
the whole three-dimensional space of solutions of Eqs.~(\oref{Feqsub3},
\oref{Geqsub3}) in the infrared region, and where 
(to 6 significant figures)
\ba
\nu \equiv \lambda\tilde Z_1 A_0 B_0^2 = 0.912771 \,, \hspace{1cm}
&\kappa = 0.769479 \,, \hspace{1cm}
&\rho = 1.96964 \mlab{asexpcond}\\
f_1 = 1  \,, \hspace{4.3cm}
&f_2 = 0.408732  \,, \hspace{1cm}
&f_3 = -0.761655  \,, \hspace{1cm}\ldots \nn\\
g_1 = \eta \equiv b_1/a_1 = 0.829602   \,, \hspace{1cm}
&g_2 = 1.31169  \,, \hspace{1cm}
&g_3 = 0.783905 \,, \hspace{1cm}\ldots \, . \nn
\ea
 
It is precisely the existence of a third independent parameter, namely
$a_1$, which allows the infrared power solution to bend over in a
logarithmic tail in a way consistent with the integral equations. 
To build a solution that is both consistent with the infrared asymptotic
expansion set up in this section and the asymptotic ultraviolet logarithmic
behaviour which will be derived in the next section, the parameter $a_1$
has to be negative, as has been inferred from the numerical results
calculated with the Runge-Kutta method and with the direct integral equation
method. If $a_1=0$ we retrieve the pure power solution
and if $a_1 > 0$ there does not seem to be a singularity-free solution
for $x \in [0,\Lambda^2]$.

As we have shown in Sect.~\ref{Sect:symmetries}, the three-dimensional
family of solutions can also be constructed once we have found one solution,
just by relying on the three distinct scale invariances (\oref{scala},
\oref{scalb}, \oref{scalt}). How these scale invariances correspond to
choices of infrared asymptotic parameters will now be elucidated.

The function scalings (\oref{scala}, \oref{scalb}) of $F(x)$, $G(x)$
correspond to similar scalings of $A_0, B_0$ in the infrared expansions
\mref{asexp2}, 
\[
\tilde A_0 = A_0/a \,, 
\hspace{1cm}
\tilde B_0 = B_0/b \,,
\]
such that condition \mref{asexpcond} remains satisfied with
$\tilde\lambda=\lambda a b^2$, and $a_1$ is left unchanged.

Less trivial is the momentum scaling invariance of the space of
solutions:
\be
\hat F(x) \equiv F(tx)
\hspace{1.5cm}
\hat G(x) \equiv G(tx) \, .
\mlab{scal2a}
\ee

Using these definitions in \mref{asexp2}, we find, after some rearrangement,
\bann
\hat F(x) &=& (t^{2\kappa} A_0) x^{2\kappa} 
\l(1 + \sum_{i=1}^{N} f_i (t^\rho a_1)^i x^{i\rho}\r) \\
\hat G(x) &=& (t^{-\kappa} B_0) x^{-\kappa} 
\l(1 + \sum_{i=1}^{N} g_i (t^\rho a_1)^i x^{i\rho}\r) \, .\nn
\eann

This shows that the infrared expansions for the momentum scaled functions
$\hat F(x), \hat G(x)$ correspond to asymptotic expansions
parametrized by
\be
\hat A_0 = t^{2\kappa} A_0 \,,
\hspace{1cm}
\hat B_0 = t^{-\kappa} B_0
\hspace{1cm}\mbox{and}\hspace{1cm}
\hat a_1 = t^\rho a_1 \,,
\mlab{scal2b}
\ee
and that the asymptotic expansions indeed obey \mref{asexp2} and the
conditions \mref{asexpcond}. 
As we expected, there is a one-to-one
correspondence between the solutions constructed from the scaling
invariances based on the symmetries of the equations, and the parameters
$A_0$, $B_0$ and $a_1$ characterizing their infrared expansions.

Let us now construct the asymptotic expansion of the running coupling
(with $\tilde Z_1=1$) using the expansions \nref{asexp2}:
\be
\lambda(x)=\lambda F(x)G^2(x) = \lambda A_0 B_0^2
\l(1 + \sum_{i=1}^{N} f_i a_1^i x^{i\rho}\r)
\l[ \l(1 + \sum_{i=1}^{N} g_i a_1^i x^{i\rho}\r) \r]^2
\ee
or (again truncating at N)
\be
\lambda(x) = \nu \l(1 + \sum_{i=1}^{N} h_i a_1^i x^{i\rho}\r) \,,
\mlab{asexp_lambda}
\ee
where
\[
h_1 = 2.65920 \,, \hspace{1cm}
h_2 = 5.37956 \,, \hspace{1cm}
h_3 = 6.97232 \,, \hspace{1cm} \ldots\,,
\]
which tells us that the running coupling only depends on the dimensionful
parameter $a_1 \equiv A_1/A_0$, and is independent of $\lambda$, $A_0$ and
$B_0$.  Furthermore, we can show from Eqs.~(\oref{scal2a}, \oref{scal2b})
that the running coupling corresponding to the parameter $\tilde a_1$, is
identical to the running coupling with parameter $a_1$ after
scaling the momentum with a factor $t=(\tilde a_1/a_1)^{1/\rho}$. This
tells us that the momentum units of $a_1$ are unambiguously related to the
physical scale of the experimentally determined running coupling.

We now introduce a momentum scale $\Omega^2$: 
\be
\Omega^2 = \frac{1}{\l(h_1|a_1|\r)^{1/\rho}} \,,
\mlab{omega}
\ee
(recall that $a_1 < 0$), such that
\be
\lambda(x) = \nu \l(1 + \sum_{i=1}^{N} (-1)^i \tilde h_i 
\l(\frac{x}{\Omega^2}\r)^{i\rho}\r) \,,
\mlab{asexp_lambda2}
\ee
where we defined
\[
\tilde h_i = \frac{h_i}{h_1^i} \,: \hspace{1cm}
\tilde h_1 = 1 \,, \hspace{1cm}
\tilde h_2 = 0.760753 \,, \hspace{1cm}
\tilde h_3 = 0.370785 \,, \hspace{1cm} \ldots \,.
\]

We will see from the numerical results that $\Omega^2$ is a good
estimate of the scale up to which the infrared asymptotic expansion
remains valid.

\section{Ultraviolet behaviour}
\label{Sect:UV-onlyghost}

We now turn to the investigation of the ultraviolet asymptotic behaviour of
the solutions. As discussed before, the numerical results show a
three-dimensional space of solutions, which has been confirmed by an
analytical study of the global symmetries of the integral equations
and by the study of the infrared asymptotic expansions of the
solutions. Except for the pure power solution, all these solutions 
bend over in a log-tail above a certain momentum scale $x$. We will now
check the consistency of such ultraviolet logarithmic solutions.

Suppose the solutions for $F(x)$ and $G(x)$, taking on the values
$F_\mu$ and $G_\mu$ at some momentum $\mu$ in the perturbative regime,
have the following ultraviolet behaviour:
\ba
F(x) &\equiv& F_\mu \, \l[\omega\log\l(\frac{x}{\mu}\r)+1\r]^\gamma
\mlab{Fuvsol-og} \\[3mm]
G(x) &\equiv& G_\mu \, \l[\omega\log\l(\frac{x}{\mu}\r)+1\r]^\delta \,.
\mlab{Guvsol-og}
\ea

We check the consistency of these ultraviolet solutions by substituting
these expressions in Eqs.~(\oref{Feqsub3}, \oref{Geqsub3}). 

The ghost equation~\nref{Geqsub3} yields, to leading log,
\be
G^{-1}_\mu \, \l[\omega\log\l(\frac{x}{\mu}\r)+1\r]^{-\delta} =
G^{-1}_\mu \, \l[\omega\log\l(\frac{\sigma}{\mu}\r)+1\r]^{-\delta} 
- \frac{9}{4} \lambda \tilde Z_1 F_\mu G_\mu 
\, \int_x^\sigma \, 
\frac{dy}{y} \, \l[\omega\log\l(\frac{y}{\mu}\r)+1\r]^{\gamma+\delta} \,. 
\mlab{GUV2-og}
\ee

After evaluating the integral we get
\ba
G^{-1}_\mu \, \l[\omega\log\l(\frac{x}{\mu}\r)+1\r]^{-\delta} &=& 
G^{-1}_\mu \, \l[\omega\log\l(\frac{\sigma}{\mu}\r)+1\r]^{-\delta}
\mlab{GUV3-og}\\ 
&& \hspace{-3cm} 
- \frac{9\lambda \tilde Z_1 F_\mu G_\mu}{4\omega (\gamma+\delta+1)}
\l\{ \l[\omega\log\l(\frac{\sigma}{\mu}\r)+1\r]^{\gamma+\delta+1}
- \l[\omega\log\l(\frac{x}{\mu}\r)+1\r]^{\gamma+\delta+1} \r\} \,. \nn 
\ea

Matching the index of the leading powers of logarithms in
\mref{GUV3-og} one finds the consistency condition:
\be
\gamma + 2\delta = -1
\mlab{uvcond1-og}
\ee
and, equating the leading log coefficients in \mref{GUV3-og}, using
\mref{uvcond1-og}, we get
\be
\lambda \tilde Z_1 F_\mu G^2_\mu = \frac{2\omega}{9} (\gamma+1) \,.
\mlab{uvcond2'-og}
\ee

Substituting the solutions Eqs.~(\oref{Fuvsol-og}, \oref{Guvsol-og})
in the gluon equation \mref{Feqsub3} and keeping only the leading
log terms, we find
\be
F^{-1}_\mu \, \l[\omega\log\l(\frac{x}{\mu}\r)+1\r]^{-\gamma}
= F^{-1}_\mu \, \l[\omega\log\l(\frac{\sigma}{\mu}\r)+1\r]^{-\gamma}
+ \lambda \tilde Z_1 G^2_\mu \,
\int_x^{\sigma} \, \frac{dy}{2y} \,
\l[\omega\log\l(\frac{y}{\mu}\r)+1\r]^{2\delta} \,.
\ee

After performing the integrals, we find
\ba
F^{-1}_\mu  \, \l[\omega\log\l(\frac{x}{\mu}\r)+1\r]^{-\gamma}
&=& F^{-1}_\mu  \, \l[\omega\log\l(\frac{\sigma}{\mu}\r)+1\r]^{-\gamma}
\mlab{FUV2-og} \\
&& \hspace{-2cm} + \frac{\lambda \tilde Z_1 G^2_\mu}{2\omega(2\delta+1)} 
\l\{ \l[\omega\log\l(\frac{\sigma}{\mu}\r)+1\r]^{2\delta+1}
- \l[\omega\log\l(\frac{x}{\mu}\r)+1\r]^{2\delta+1} \r\} \,. \nn
\ea

Consistency of the exponents on both sides of the equation is
automatically guaranteed by \mref{uvcond1-og}. Then, equating the
coefficients of the leading log contributions of \mref{FUV2-og}, and
substituting \mref{uvcond1-og}, we obtain
\be
\lambda \tilde Z_1 F_\mu G^2_\mu = 2\omega\gamma \,.
\mlab{uvcond3-og}
\ee

From Eqs.~(\oref{uvcond1-og}, \oref{uvcond2'-og}, \oref{uvcond3-og}) we
then find 
\be 
\gamma = \frac{1}{8} \hspace{1.5cm} \delta=-\frac{9}{16}
\ee
and the equivalent conditions Eqs.~(\oref{uvcond2'-og}, \oref{uvcond3-og})
yield
\be 
\omega = 4 \lambda \tilde Z_1 F_\mu G^2_\mu \, .
\mlab{uvcond2''-og}
\ee

Thus, the ultraviolet solutions for $F(x), G(x)$ can be written as
\ba
F(x) &=& F_\mu \, \l[4 \lambda \tilde Z_1 F_\mu G^2_\mu
\log\l(\frac{x}{\mu}\r)+1\r]^{1/8} \mlab{Fuvsol1-og} \\
G(x) &=& G_\mu \, \l[4 \lambda \tilde Z_1 F_\mu G^2_\mu
\log\l(\frac{x}{\mu}\r)+1\r]^{-9/16} \mlab{Guvsol1-og}
\ea
and the renormalization group invariant running coupling is given by
\be
\lambda(x) = \lambda F(x) G^2(x) =
\frac{1}{ 4 \tilde Z_1 \log\l(\frac{x}{\mu}\r)
+\frac{1}{\lambda F_\mu \, G^2_\mu}} \,.
\ee

We can rewrite this in the form 
\be
\lambda(x) = \frac{1}{ \beta_0 \log\l(\frac{x}{\Lambda^2_{QCD}}\r)} \,
 , \mlab{rc-uv}
\ee
where $\beta_0=4$, and the QCD scale is given by  
\be
\Lambda^2_{QCD} 
= \mu \exp\l( -\frac{1}{4 \lambda F_\mu G^2_\mu }\r) \,,
\mlab{LambdaQCD}
\ee
if $\tilde Z_1 = 1$. We see that fixing $\lambda F_\mu G^2_\mu$ at a
scale $\mu$, in the perturbative regime, 
indeed amounts to a definition the value of $\Lambda_{QCD}$.

The leading-log coefficient is $\beta_0=4$,  
but this is not
in agreement with perturbation theory, where $\beta_0=11$. However, the
reason for this is obvious, as we only considered the ghost loop and
discarded the gluon loop in the gluon equation.

\section{Numerical method}

Knowing the infrared and ultraviolet asymptotic behaviours of the coupled
equations Eqs.~(\oref{Feqsub3}, \oref{Geqsub3}), we now go on to solve the
equations numerically in order to see if we can find consistent
solutions over the whole momentum range, connecting both asymptotic
regions, hopefully giving us more insight into the transition from the regime
of asymptotic freedom to the state of confinement. We first give a short
overview of the numerical method we used.

We use a numerical method developed by one of us for the study of dynamical
fermion mass generation in QED$_4$\cite{Bloch95b}. This method directly
solves the coupled integral equations by an iterative numerical scheme. We
also checked the results, found with the integral equation method, with a
Runge-Kutta method applied to the set of differential equations derived
from the integral equations.

We now give an outline of the main features of the integral equation
method. Unlike most other methods used thus far, we replaced the widely
used discretization of the unknown functions by smooth polynomial
approximations, introducing Chebyshev expansions for the gluon and ghost
form factors $F(x)$ and $G(x)$ and using the logarithm of momentum squared
as variable. To improve the accuracy of the Chebyshev approximations we
first extract the infrared power behaviours of the form factors, although
this only has a minor influence. The form factors are approximated by
\ba
F(x) \equiv A x^{2\kappa} \l[\frac{a_0}{2} + \sum_{j=1}^{N-1} a_j
T_j(s(x))\r] \mlab{Fexp}\\
G(x) \equiv B x^{-\kappa} \l[\frac{b_0}{2} + \sum_{j=1}^{N-1} b_j
T_j(s(x))\r]
\mlab{Gexp}
\ea
with
\be
s(x) \equiv \frac{\logten(x/\Lambda\epsilon)}{\logten(\Lambda/\epsilon)} \,,
\mlab{912}
\ee
and where $\Lambda$ is the ultraviolet cutoff, and $\epsilon$ is the
infrared cutoff, only needed for numerical purposes.  We require both
integral equations to be satisfied at $N$ fixed external momenta, in order
to determine the $2N$ Chebyshev coefficients $a_i, b_i$. Using smooth
expansions has the advantage of allowing us an absolute freedom in the
choice of quadrature rules used to compute the various integrals
numerically. This is required if we want to achieve a high accuracy in our
results. The integration region is first split into an analytical integral
over $[0, \epsilon^2]$ and a numerical integral over $[\epsilon^2,
\Lambda^2]$. The integral over $[0, \epsilon^2]$ is computed analytically
from the asymptotic infrared behaviour discussed in
Sect.~\ref{Sect:IR}. This is needed as the infrared part of the integral is
highly non-negligible, especially in the case of the gluon equation.  For
an efficient choice of quadrature rule we split the numerical integral into
three regions, these are $[\epsilon^2, \min(x,\sigma)]$,
$[\min(x,\sigma),\max(x,\sigma)]$ and $[\max(x,\sigma), \Lambda^2]$,
where $x$ is the external momentum and $\sigma$ is the subtraction
point. The splitting of the region of numerical integration into three
subregions is needed as the integrands are not smooth at the boundaries of
these regions and too much accuracy is lost if one uses quadrature rules
spanning these boundaries. A sensible choice of quadrature rule on each
integration region is for instance a composite 4-points Gaussian
integration rule, where the composite rules are delimited by the region
boundaries and the values of the external momenta at which we require the
integral equation to be satisfied. This setup will yield $2N$ coupled,
non-linear, algebraic equations for the $2N$ Chebyshev coefficients $a_j$
and $b_j$. In traditional Dyson-Schwinger studies, the unknowns are usually
determined by what is often called the {\it natural iteration method},
where the current approximation to the unknowns is used in the integrals of
the right-hand side of the equations in order to provide a new
approximation to the unknowns used in the left-hand side of the
equations. This iteration method however is not necessarily convergent, and
when it is convergent it often converges very slowly, as has been shown in
Ref.\cite{Bloch95b}. This slow rate of convergence is not only inefficient, but
more importantly it makes it very difficult to get a reliable estimate of
the accuracy of the solution. For this reason, our numerical method uses
the Newton method to solve sets of non-linear equations. This method uses
the derivatives of the equations with respect to the unknowns to speed up
the convergence. If the starting guess to the unknown coefficients is close
enough to the solution, the convergence rate is even
quadratic. Let us symbolically rewrite the coupled
functional equations as follows: 
\bann
f(x)[F,G] &=& 0 \\
g(x)[F,G] &=& 0 \,,
\eann
where $f$ and $g$ are equivalent to the
Eqs.~(\oref{Feqsub3}, \oref{Geqsub3}) and $x \in [0,\Lambda^2]$. For the
numerical solution, we require this equation to be satisfied at the
external momenta $x_i$, the functions $F$ and $G$ are expanded as Chebyshev
polynomials with coefficients $a_j$ and $b_j$, and the integrals are
approximated by a suitable quadrature rule. The equations then become
\bann
\tilde f(x_i)[a_j,b_j] &=& 0 \\
\tilde g(x_i)[a_j,b_j] &=& 0 \,,
\eann
where $i,j=0 \ldots N-1$ and $\tilde f$ and $\tilde g$ are the numerical
approximations to $f$ and $g$ when the integrals are replaced with
quadrature rules.

The Newton method will yield successive approximations to the
solutions, given by
\bann
a_j^{n+1} &=& a_j^n - \Delta a_j^{n+1} \\
b_j^{n+1} &=& b_j^n - \Delta b_j^{n+1} \,,
\eann
and the $(n+1)$-th
improvements $\Delta a_j^{n+1}$, $\Delta b_j^{n+1}$ are given by the
solutions of the $2N$x$2N$ set of linear equations
\bann
\frac{\delta \tilde f^n(x_i)}{\delta a_j}\Delta a_j^{n+1} 
+ \frac{\delta \tilde f^n(x_i)}{\delta b_j}\Delta b_j^{n+1} = 0 \\ 
\frac{\delta \tilde g^n(x_i)}{\delta a_j}\Delta a_j^{n+1} 
+ \frac{\delta \tilde g^n(x_i)}{\delta b_j}\Delta b_j^{n+1} = 0 \,,
\eann
where the equations are taken at the $N$ external momenta $x_i$ and
each equation includes implicit summations over $j$.

The total accuracy depends on the combination of the accuracies of the
Chebyshev expansion and of the quadrature rule and on the convergence
criterion of the Newton iteration. We will see later, after comparing the
results of this method with those obtained with the Runge-Kutta method,
that we can achieve a very high accuracy over quite a broad momentum range.

Using this method, we performed a meticulous study of the equations
Eqs.~(\oref{Feqsub3}, \oref{Geqsub3}). We note that, for a fixed value of
$\lambda$, the equations have two free parameters, for instance $F(\sigma)$
and $G(\sigma)$ (restricted by \mref{nu}, $\lambda F(\sigma)G^2(\sigma) \le
0.912771$). Furthermore, as shown in Sect.~\ref{Sect:symmetries}, a scaling
of $\lambda$ can always be absorbed in a redefinition of the unknown
functions $F(x)$ and $G(x)$, such that knowing the solution space for
one value of $\lambda$, we can build the solutions for any arbitrary value
of $\lambda$ in a straightforward way. Moreover, such scalings of $\lambda$
leave the running coupling $\lambda F(x) G^2(x)$ unchanged.

In practice, we choose an alternative pair of parameters, $F(\sigma)$ and $A$,
where $A$ is the leading infrared gluon coefficient defined in
\mref{FIR}. The choice of these two parameters is suggested by the
numerical solution method. Using \mref{nu}, the value of $A$ also
determines the leading infrared ghost coefficient $B$, and allows us to
compute a quite accurate analytical approximation to the infrared part of
the integral over $[0,\epsilon^2]$, if $\epsilon^2$ is sufficiently
small. The choice of $F(\sigma)$ as second parameter can be viewed as a
measure of the deviation from the pure power behaviour at momentum $\sigma$.
We have taken the subtraction scale to be
$\sigma=1$ and varied both parameters $A$ and $F(1)=F_1$ for a fixed
value of $\lambda=1$ and $\tilde Z_1=1$.

\section{Results}

We vary both parameters $A$ and $F_1$ to scan the two parameter space of
solutions, keeping $\lambda=1$ fixed. As expected we retrieve the scaling
invariances discussed in the previous sections.

\begin{figure}
\begin{center}
\input{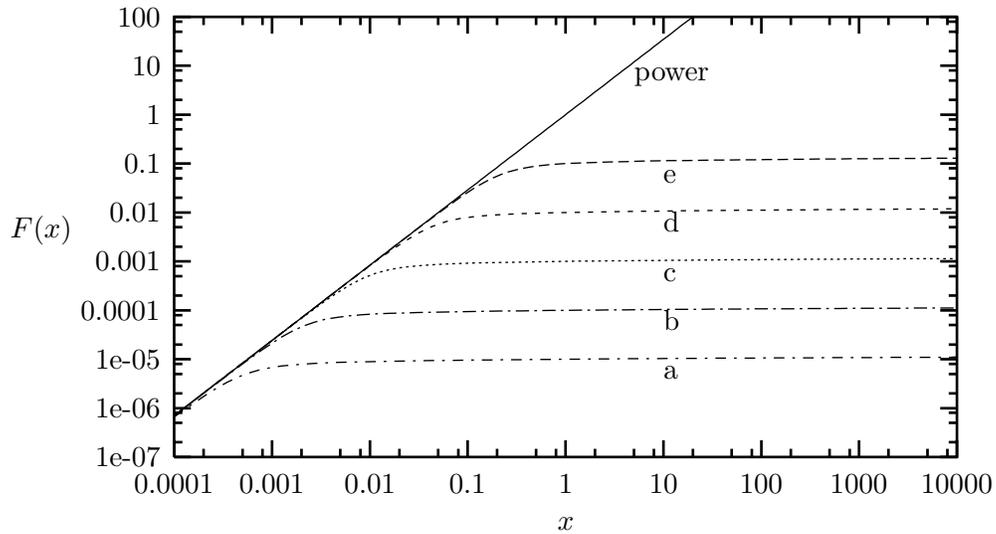}
\input{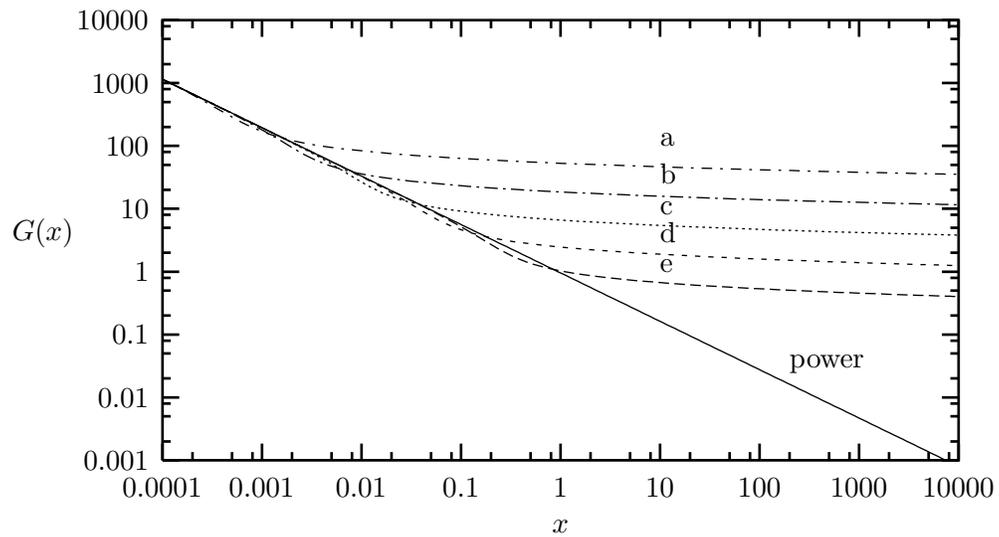}
\end{center}
\vspace{-1cm} \caption{Gluon and ghost form factors $F(x)$ and $G(x)$
versus momentum $x$ (on log-log plot), for $\lambda=1$, $A=1$ and 
$F_1=$ $10^{-5}$(a), $10^{-4}$(b), $10^{-3}$(c), 0.01(d) and 0.1(e).}
\label{Fig:plot2}
\end{figure}

If we plot the solutions for $F(x)$ and $G(x)$ for various sets
($A$,  $F_1$), as in Fig.~\ref{Fig:plot2}, we can check that every
solution can be transformed into another one by a unique
transformation (t,r) corresponding to a momentum scaling $tx$ and a
function scaling $rF(x)$, $G(x)/\sqrt{r}$. 
The numerical results clearly show the expected power behaviour in the
infrared region and the logarithmic behaviour in the ultraviolet
region. The value of the exponents and of the coefficients in front of the
power in the infrared region is completely consistent with the analytical
treatment of Sect.~\ref{Sect:IR}, which was also used to compute the
infrared part $[0,\epsilon^2]$ of the integrals analytically. As can be
seen from the plots, the gluon form factor, which starts off as a power
with a given coefficient $A$, will bend over at some cross-over point
$\tilde x$, such that the further logarithmic behaviour of the function
consistently leads to a value $F_1$ at the subtraction scale
$\sigma=1$. The logarithmic behaviour of $F(x)$ and $G(x)$ also satisfies
the ultraviolet leading log behaviour analyzed in
Sect.~\ref{Sect:UV-onlyghost}. It is remarkable that both asymptotic
regimes, infrared and ultraviolet, seem to connect onto each other at some
momentum $\tilde x$, with scarcely any intermediate regime.

\begin{figure}
\begin{center}
\input{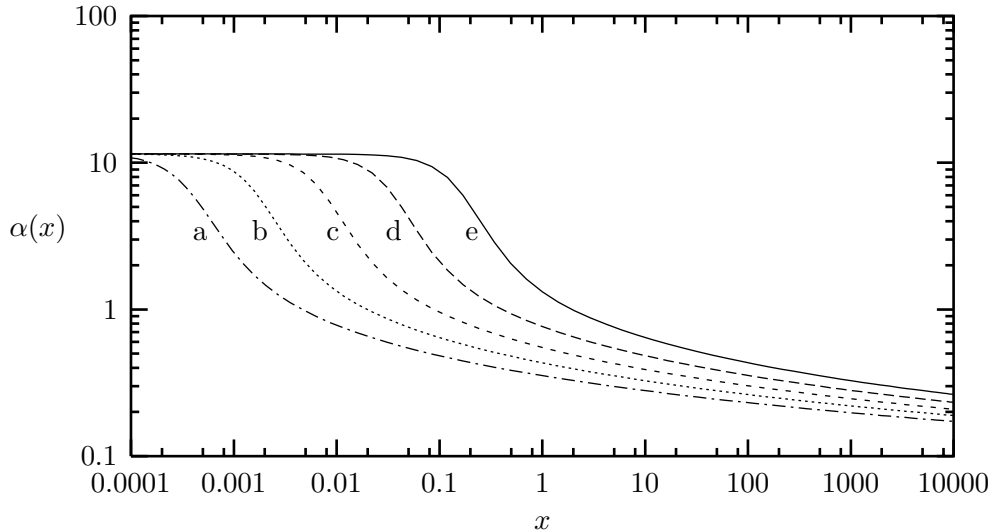}
\end{center}
\vspace{-1cm}
\caption{Running coupling $\alpha(x)=\alpha F(x) G^2(x)$ versus momentum
$x$ (on log-log plot), for $A=1$ and $F_1=$ $10^{-5}$(a),
$10^{-4}$(b), $10^{-3}$(c), 0.01(d) and 0.1(e).}
\label{Fig:plot1}
\end{figure}

If we look at the running coupling we see that all the solutions are
just translations of each other when plotted on a logarithmic momentum
scale, as is illustrated in Fig.~\ref{Fig:plot1}. This corresponds to
the invariance of the space of solutions with respect to scaling of
momentum. It also shows the physical equivalence of all solutions as
such a transformation can always been absorbed into a redefinition of
momentum units.

It is also interesting to compare the numerical results with the analytic
asymptotic calculations in order to investigate in which momentum regions
the asymptotic solutions are valid. As example we consider the case $A=1$
and $F_1=0.1$. To compute the infrared asymptotic expansion we need to know
the value of the infrared parameter $a_1$ in \mref{asexp2}. We used the
Runge-Kutta method, which will be described in Sect.~\ref{Sect:RK},
in order to determine the value of $a_1$ yielding a value of $F_1(1)=0.1$
for the gluon form factor, with $A=1$. For this specific case, the value is
$a_1 \approx -10.27685$ or $\Omega^2\approx 0.186475$ (from
\mref{omega}). The infrared asymptotic expansion,
derived in Sect.~\ref{Sect:irasexp}, is calculated from
\mref{asexp_lambda2} and truncated after four terms:
\[
\alpha(x) \stackrel{\rm ir}{\sim}
4\pi\nu
\l[1 - \l(\frac{x}{\Omega^2}\r)^\rho 
+ 0.760753\l(\frac{x}{\Omega^2}\r)^{2\rho} 
- 0.370785\l(\frac{x}{\Omega^2}\r)^{3\rho}\r] \, .
\]

The ultraviolet asymptotic behaviour, derived in
Sect.~\ref{Sect:UV-onlyghost}, is described by \mref{rc-uv}:
\[
\alpha(x) \stackrel{\rm uv}{\sim}
\frac{4\pi}{4\log\l(\frac{x}{\Lambda_{QCD}^2}\r)} \,,
\]
where we use \mref{LambdaQCD},
\[
\Lambda^2_{QCD} = \mu \exp\l(-\frac{1}{4\lambda(\mu)}\r) \,,
\]
to compute the value of $\Lambda_{QCD}$ for the case under
consideration. We choose $\mu$ in the perturbative regime, for example
$\mu=1032.15$, where the numerical results yield $\lambda(\mu)\equiv\lambda
F(\mu) G^2(\mu) \approx 0.0259676$, and find
\[
\Lambda^2_{QCD} = 0.06802 \,,
\]
still in arbitrary units.

\begin{figure}
\begin{center}
\input{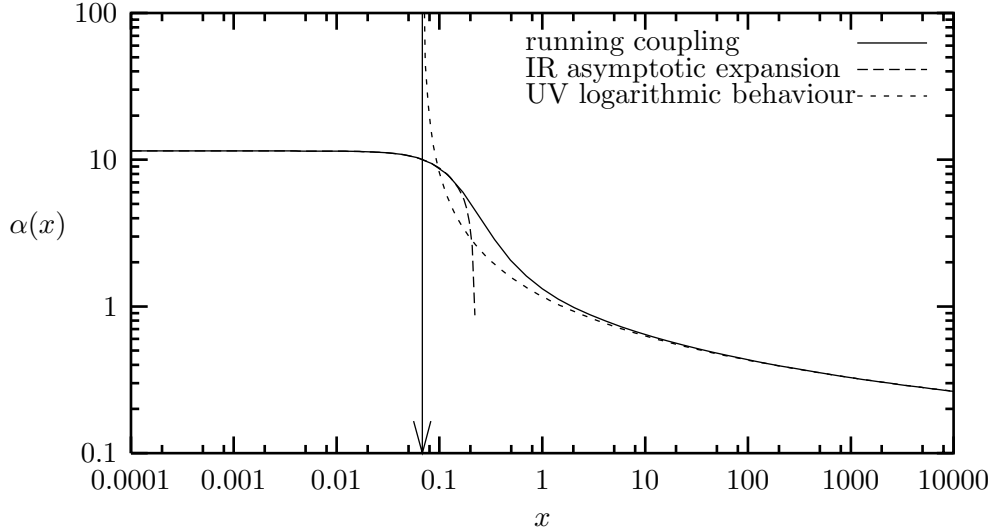}
\end{center}
\vspace{-1cm}
\caption{Running coupling $\alpha F(x) G^2(x)$ versus momentum $x$ (on
log-log plot), for parameter values $A=1$ and $F_1=0.1$ together with its
infrared asymptotic expansion and its ultraviolet asymptotic behaviour.}
\label{Fig:plot3}
\end{figure}

In Fig.~\ref{Fig:plot3} we plot the running coupling versus momentum,
together with its infrared asymptotic expansion and the ultraviolet
asymptotic behaviour. 
The agreement between the analytical and numerical
results is extremely good, and it can be seen that both asymptotic
behaviours flow into each other, almost without any intermediate
regime. 
The vertical line in
Fig.~\ref{Fig:plot3} situates the scale of $\Lambda_{QCD}^2$. We see that
$\Lambda_{QCD}^2$ lies in the momentum regime where the infrared asymptotic
expansion has already taken over from the logarithmic behaviour, and where
the running coupling has become almost constant. 
Furthermore, the infrared scale $\Omega^2\approx 0.186$ seems to be a good
measure to delimit the infrared region where the asymptotic expansion is
valid.

We can even give a numerical relation between $\Omega^2$ and
$\Lambda_{QCD}^2$ (where the latter is computed from leading log
only), namely 
\[
\frac{\Lambda_{QCD}^2}{\Omega^2} \approx 2.74 \,,
\]
and the ratio is the same for all solutions of the
equations~(\oref{Feqsub3}, \oref{Geqsub3}). The simple relation between
$\Omega$ and $\Lambda_{QCD}$ is a consequence of the symmetries of the {\it
ghost-loop-only} truncation. If we include the gluon loop in the same
truncation scheme, the asymptotic expansion of the running coupling will no
longer depend on $a_1$ alone, but on the other parameters of the infrared
expansions as well. Hence an ultraviolet renormalization, leading to a
specific value of $\Lambda_{QCD}^2$, will correspond to a family of running
couplings, all having slightly different behaviours in the intermediate
regime, and there will be an ambiguity in the determination of the
non-perturbative running coupling.

The units in which $\Lambda_{QCD}$ is expressed are still arbitrary,
as we still have to compare the numerical results with experimental
data. In the truncation under consideration, comparison with experimental
results is only interesting to check our methodology. The numbers which
will come out of the analysis are not to be taken too seriously, as the
leading-log $\beta$-coefficient in the {\it ghost-loop-only} case is 4,
while in the pure gauge theory it ought to be 11, and in the presence of
$n_f$ flavours of fermions it should become $(33-2n_f)/3$, as is known from
perturbation theory. This means that the ultraviolet running of the
coupling will be too slow in the case we are considering, and the value of
$\Lambda_{QCD}$ will come out far too low. If we use the PDG\cite{PDG}
value $\alpha_s(M_Z=91.187~\GeV)=0.118$, we find $\Lambda_{QCD}^2=
2.277\times 10^{-8}~\GeV^2$ (using \mref{LambdaQCD}).  Thus, if we want to
map the numerical results discussed above (expressed in arbitrary momentum
units {\it amu}) to physical reality we have to set $1(amu)^2 =
3.348\times 10^{-7}~\GeV^2$ and the mass of the Z-boson will be at
$2.484\times 10^{10}~(amu)^2$. Of course all this is only hypothetical as
the $\beta$-coefficient of the physical theory should be 25/3 in the
presence of 4 fermion flavours, and this would yield a value
$\Lambda_{QCD}^2=0.02343~\GeV^2$. We also note that
the freedom of renormalization allows us to choose the renormalized
quantities, as is usually done in perturbation theory,
i.e. $\alpha_s(M_Z=91.187~\GeV)=0.118$, $F(M_Z)=G(M_Z)=1$. This is possible,
as all solutions of the three-dimensional space of solutions are physically
equivalent. We will see later that this is not as obvious as it seems, and
that it depends on the truncation scheme. Including the gluon-loop (with
$Z_1=1$) in the gluon equation, either with a bare triple gluon vertex or
even with a Ball-Chiu vertex, will destroy this invariance and different
solutions will correspond to couplings running in different ways.

\section{Starting guess}

The Newton method, which is at the core of our numerical method, is a
quadratically convergent iterative method, if the initial approximations to
the unknown functions are {\it sufficiently close} to the exact
solutions. The meaning of sufficiently close depends however entirely on
the kernel of the integral equation. We observed that for the coupled
gluon-ghost equations, the starting guess must not be too remote from the
exact solution, if the method is to converge. This is in contrast with
previous work on chiral symmetry breaking in QED and on the Mandelstam
approximation to the gluon propagator in QCD, where the method was
extremely insensitive to the starting guess. 

It turns out that in the case of the coupled gluon-ghost equations, the
starting guesses have to be chosen quite sensibly, especially in the
asymptotic regions. In practice, we used the analytic asymptotic solutions to
build good enough starting guesses for the form factors. 

\begin{figure}
\begin{center}
\input{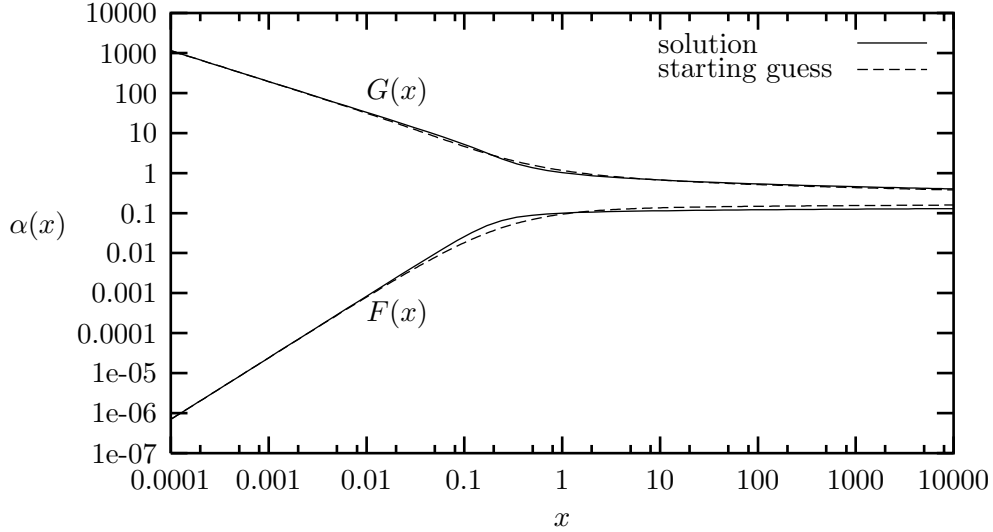}
\end{center}
\vspace{-1cm} 
\caption{Comparison of the solutions for $F(x)$ and $G(x)$ with their
starting guesses used in the iterative Newton method, for $\lambda=1$,
$A=1$ and $F_1=0.1$.}
\label{Fig:plot6}
\end{figure}

Given the parameters $A$ and $F_1$, the leading order infrared ghost
coefficient is
\[
B = \sqrt{\frac{\nu}{\lambda \tilde Z_1 A}} \,,
\]
and we define $\tilde x$ as
\[
\tilde x = \l(\frac{F_1}{A}\r)^{1/2\kappa} \,,
\]
which can be seen as a crude approximation to the bend-over point.
A possible construction is:
\bann
F(x) &=& A\l[\frac{x}{\frac{x}{\tilde x} + 1}\r]^{2\kappa}
\l[4\nu \log\l(\frac{x}{\tilde x} + 1\r)+1\r]^{1/8} \\
G(x) &=& B\l[\frac{x}{\frac{x}{\tilde x} + 1}\r]^{-\kappa} 
\l[4\nu \log\l(\frac{x}{\tilde x} + 1\r)+1\r]^{-9/16} \, ,
\eann
which has the correct leading infrared asymptotic behaviour for $F(x)$ and
$G(x)$ and agrees well with their leading ultraviolet logarithmic
behaviour, as is illustrated in Fig.~\ref{Fig:plot6}.

\begin{figure}
\begin{center}
\input{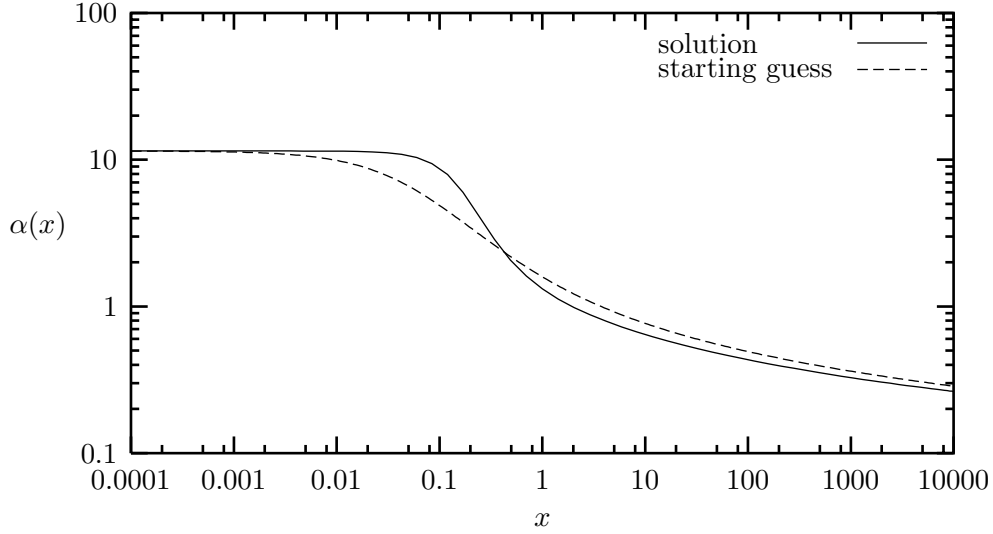}
\end{center}
\vspace{-1cm} \caption{Comparison of the solution for the running coupling
$\alpha(x)$ with its starting guess, for $\lambda=1$, $A=1$ and $F_1=0.1$.}
\label{Fig:plot7}
\end{figure}

Although it seems that the starting guesses $F(x)$ and $G(x)$ are extremely
close to the eventual numerical solutions, we see that the Newton method
does alter the running coupling $\alpha(x)=4\pi\lambda F(x) G^2(x)$
substantially while converging to the solution, as is shown in
Fig.~\ref{Fig:plot7}.

\section{Runge-Kutta method}
\label{Sect:RK}

Rewrite the equations (\oref{Feqsub3}, \oref{Geqsub3}), for $\sigma=1$, as 
\be
F^{-1}(x)=\eta +\lambda
\biggl[ 
\frac{G(x)}{x^2}\int_0^xdy
\left( \frac{3y}{2}-\frac{y^2}{x}\right)G(y)
+\int_x^1\frac{dy}{2y}G^2(y)\biggr] 
\mlab{F4}
\ee
and
\be
\mlab{G4}
G^{-1}(x)=\zeta 
-\frac{9}{4}\lambda
\biggl[ 
\frac{F(x)}{x^2}\int_0^xdy y G(y) + 
\int^1_x \frac{dy}{y}F(y)G(y)\biggr] \, ,
\ee
where $\eta$ and $\zeta$ are constants. 
As discussed before we can choose $\lambda=1$, since an arbitrary value of
$\lambda$ can be recovered by applying an appropriate scaling to the form
factors $F(x)$ and $G(x)$.

Let us rewrite the above equations in the form 
\be
{F}^{-1}(x)=\eta + \frac{3}{2}
G(x)K(x) - G(x)L(x)
+\frac{1}{2}\int_x^1\frac{dy}{y}{G}^2(y)
\mlab{F5}
\ee
and
\be
\mlab{G5}
{G}^{-1}(x)=\zeta 
-\frac{9}{4}
{F}(x)K(x)
-\frac{9}{4}
\int^1_x \frac{dy}{y}{F}(y){G}(y) \, ,
\ee
where
\be
K(x)=\frac{1}{x^2}\int_0^xdy y{G}(y)
\mlab{K5}
\ee
and
\be
\mlab{L5}
L(x)=\frac{1}{x^3}\int_0^xdy y^2{G}(y)\, .
\ee

On differentiating the above four equations, we obtain 
\bann
\dot{F} &=&F^2[-\mfrac{3}{2}\dot{G} K-\mfrac{3}{2}G\dot{K} 
+\dot{G} L+G\dot{L} ]+\half F^2G^2\\
  \dot{G} &=&\mfrac{9}{4}G^2[\dot{F}  K+F\dot{K} ]-\mfrac{9}{4}FG^3 \\
 \dot{K} &=&G-2K\\
 \dot{L} &=&G-3L\, , 
\eann
where $\dot{F}=\frac{dF}{dt}=x\frac{dF}{dx}$ etc., with $t=\log x$. 
After a little algebra, we can throw the first two of these equations into the form
\ba
\dot{F} &=&3F(X-Y)-FZ(\mfrac{3}{2}X-Y)\nonumber \\
\dot{G} &=&ZG\, , \mlab{RKnew}
\ea
where 
\ba
 X &=& FGK \nonumber \\
 Y &=& FGL \nonumber \\
 Z &=& \frac{X(3X-3Y-2)}{\mfrac{4}{9}+X(\mfrac{3}{2}X-Y)}\, .
\mlab{XYZ}
\ea

This system is suitable for an application of the Runge-Kutta method; but
we must first address the question of the existence and multiplicity of the
solutions, first of the differential system, and then of the integral
equations (\oref{F5}, \oref{G5}) from which they were derived.  We already
know an (exact) solution, namely
\ba
 F(x) &=& A x^{2\kappa} \nonumber \\
 G(x) &=& B x^{-\kappa}\, , \mlab{FGk}
\ea
where $\kappa$ was defined in \mref{kappa2}, and where (compare
\mref{AB2-2} with $\lambda=1$)
\be
\mlab{g0}
B=\frac{2}{3\sqrt{A}}
\left( \frac{1}{\kappa}-\frac{1}{2-\kappa}\right)^{-\frac{1}{2}}
\approx \frac{0.955}{\sqrt{A}} \, .
\ee

{\em A priori} we would expect there to be {\em four} free parameters
for the differential system, corresponding say to the values of
$F(1)$, $G(1)$, $K(1)$ and $L(1)$, from which the differential equations 
could step-by-step be integrated, for example by the Runge-Kutta
method. In general, solutions of the differential system would
not satisfy the requirements $x^2K(x)\rightarrow 0$ and
$x^3L(x)\rightarrow 0$ as $x\rightarrow 0$. In fact the lower limits
of the integrals in Eqs.(\oref{K5}, \oref{L5}) would be incorrectly replaced
by nonzero constants. Imposing the requisite boundary conditions at $x=0$, we
expect to reduce the number of arbitrary constants in the general
solution from four to two. Since there is a scaling invariance that
leaves $FG^2$ unchanged, this means that, after we have removed this
trivial degree of freedom by fixing $A$ in \mref{FGk}, we should still
have one non-trivial free parameter. Where is it?

In Sect.~\ref{Sect:irasexp} we have seen that we can construct the
following infrared asymptotic expansion, \mref{asexp2}, for the general
solutions $F(x)$ and $G(x)$:
\ba
\mlab{F+G}
F(x) &=& A_0 x^{2\kappa} \l(1 + \sum_{i=1}^{N} f_i a_1^i x^{i\rho}\r) \\
G(x) &=& B_0 x^{-\kappa} \l(1 + \sum_{i=1}^{N} g_i a_1^i x^{i\rho}\r) \,. \nn
\ea
We expect these series to have zero radius of convergence, 
so they have been truncated in the anticipation that they are asymptotic 
series --- that is, for small values of $x$, there will be an optimal  
truncation point, $N$, for which the finite series is a good approximation. 
From this expansion we see that, besides the parameters $A_0$ and $B_0$, 
there is one more free parameter, $a_1$.

On substituting the series for $G$ in the definitions Eqs.~(\oref{K5},
\oref{L5}), we find
\ba
\mlab{K+L}
K(x)&=& B_0 x^{-\kappa} \l(\frac{1}{-\kappa+2} + \sum_{i=1}^{N}
\frac{g_i a_1^i x^{i\rho}}{-\kappa +i\rho+2}\r) \\
L(x)&=& B_0 x^{-\kappa} \l(\frac{1}{-\kappa+3} + \sum_{i=1}^{N}
\frac{g_i a_1^i x^{i\rho}}{-\kappa +i\rho+3}\r) \nn\, .
\ea

The knowledge of the infrared asymptotic expansions for $F(x)$,
$G(x)$, $K(x)$ and $L(x)$ allows us to use a Runge-Kutta method, 
starting from a momentum point deep in the infrared region and
building the solution for increasing momenta. The Runge-Kutta method
was run using the {\sl NDSolve} routine of Mathematica 3.0. The
problem is solved as a function of $t=\log{x}$ and as the starting
point, the IR series Eqs.~(\oref{F+G}, \oref{K+L}) are evaluated at
$x=0.0001$ with $N=8$, using the coefficients $f_j$ and $g_j$ which
are calculated with Mathematica as well. The Runge-Kutta routine is
run with 25 digit precision and 10,000 steps from $x=10^{-4}$ to
$x=10^4$ for various values of $a_1<0$. The results produced by
this method agree extremely well with those found with the
direct integral equation method, as we will see in the next section.

\section{Comparing the Runge-Kutta and the direct method}

It is interesting to compare the two numerical methods used to solve the
coupled set of integral equations. The Runge-Kutta method is a local
method, which computes the function values at each point using the function
values at neighbouring points, starting from a momentum value deep in the
infrared region and the asymptotic expansion at that point, while the
direct integral equation method is a global method, the complete momentum
range being solved simultaneously. Each method employs a different set of
parameters. For a given $\lambda$, the Runge-Kutta method uses the
infrared coefficients $A_0$ and $a_1$, while the direct method uses $A_0$
and $F_1$. To compare results, we first have to determine the parameter
sets corresponding to the same solution in the three-dimensional space of
solutions. We run the Runge-Kutta method with $\lambda=1$, $A_0=1$, and let
$a_1$ vary till we find the solution yielding $F(1)=0.1$. As mentioned
before this is found for $a_1=-10.27685$. We then compute the solutions of
the Runge-Kutta method at the N values of external momenta used in the
direct integral equation method and compare the numerical values found with
both methods, using the maximum norm. For $N=81$, we find
\bann
\norm{F^{\rm dir}-F^{\rm RK}} \equiv \max_{i=0}^{N-1} |F^{\rm
dir}(x_i)-F^{\rm RK}(x_i)| &=& 5.7 \times 10^{-5} \\
\norm{G^{\rm dir}-G^{\rm RK}} \equiv \max_{i=0}^{N-1} |G^{\rm
dir}(x_i)-G^{\rm RK}(x_i)| &=& 6.0 \times 10^{-5} \,.
\eann

The agreement between these two very different numerical solution methods
by far surpasses our initial expectations. Especially for the direct method
it was hoped that the accuracy would be between $1/100$ and
$1/1000$. However, the above mentioned numbers show that also this method
achieves an even better accuracy. 

The Newton iteration of the direct method requires about 4 iterations to
converge and the program needs approximatively 19 sec. real time to run on
a Linux operated Pentium 200MHz PC. The Runge-Kutta method runs in
approximatively 9 sec. using the Mathematica 3.0 routine {\it NDSolve} on
the same computer. The use of two different methods is extremely important,
to check the validity and accuracy of the solutions, especially in the case
where the family of solutions is quite intricate.

Although the Runge-Kutta method is faster and very accurate, it
can only be used if the integral equations can be transformed into
differential equations + boundary conditions. It also requires a very
accurate evaluation of the starting values of the functions using the
infrared asymptotic expansion. When the problem cannot be turned
into differential equations, only the direct method will be usable.

\section{Including the gluon loop}

We will now briefly discuss Eqs.~(\oref{Feqsub},
\oref{Geqsub}), i.e. the equations where both gluon loop and ghost
loop are included in the gluon equation. Although it is this specific
truncation which attracted our attention when we started the
investigation of the coupled gluon-ghost equations, our physical
expectations were better met by omitting the gluon loop. The
invariances we have discussed yielded an unambiguous running coupling,
determined by one physically relevant parameter,
$\Lambda_{QCD}$. In the following subsections, we will briefly show what
changes occur when we do include the gluon loop and why an ambiguity
occurs.

\subsection{Symmetries of the equations}

We can repeat the analysis of Sect.~\ref{Sect:symmetries} in the truncation
we are considering now. It is easy to see that the solution space will
still be invariant under scaling of momentum \nref{scalt}, i.e. when
scaling the momentum of any solution of the equations, we retrieve another
solution of the same equations. However, the two-parameter
scaling invariance (\oref{scala}, \oref{scalb}), with respect to the
functions themselves, is now reduced to a one-parameter scaling invariance
because of the additional constraint $a=b$ on the scaling factors, which
comes from adding the gluon loop. While the {\it ghost-loop-only} case was
solely a function of products $F(x)G(y)G(z)$, for various combinations of
$x, y, z$, the current truncation depends on $F(x)F(y)F(z)$ as well
as on $F(x)G(y)G(z)$. The fact that the three-dimensional space of
solutions has lost part of its symmetry is important, as it means that
$\lambda F(x) G^2(x)$ is not unique, even after an appropriate scaling of
momentum. Globally we can say that $\lambda F(x)G^2(x)$ is no longer invariant,
because of the admixture of $\lambda F^3(x)$ terms.

\subsection{Infrared behaviour}

Because the ghost equation remains unchanged, it is easy to see that the
leading infrared behaviour in this case will be the same as in the {\it
ghost-loop-only} case. The additional gluon loop in the gluon equation only
yields higher order corrections. The asymptotic expansion set up in
Sect.~\ref{Sect:irasexp} is still generated in this case, but at some
higher order it will have to be supplemented by other higher order series,
which will be related to the leading asymptotic series. We also note
that the power solution will not be an exact solution of the equations any
more, although it remains the correct leading infrared asymptotic
behaviour.

\subsection{Ultraviolet behaviour}
\label{Sect:UV}

We will show that the leading log ultraviolet behaviour of the running
coupling still has the $1/\log{x}$ behaviour, as expected from perturbation
theory, but that the $\beta$-coefficient is different from the perturbative
one. This discrepancy is a bit surprising, since one expects the
perturbative result to be contained in the ghost and gluon equations
considered. The reason why this happens is that, for some reason, the pure
perturbative result does not consistently solve the non-perturbative
equations.

As in Sect.~\ref{Sect:UV-onlyghost}, we try the following ultraviolet
solutions for $F(x)$ and $G(x)$, taking on the values $F_\mu$ and $G_\mu$
at some momentum $\mu$ in the perturbative regime:
\ba
F(x) &\equiv& 
F_\mu \, \l[\omega\log\l(\frac{x}{\mu}\r)+1\r]^\gamma \mlab{Fuvsol} \\
G(x) &\equiv& 
G_\mu \, \l[\omega\log\l(\frac{x}{\mu}\r)+1\r]^\delta \mlab{Guvsol} \,.
\ea

We check the consistency of these ultraviolet solutions by substituting
the expressions in Eqs.~(\oref{Feqsub}, \oref{Geqsub}). For the ghost
equation \mref{Geqsub} the treatment is identical to that of
Sect.~\ref{Sect:UV-onlyghost} and we again have
\be
\gamma + 2\delta = -1 \,,
\mlab{uvcond1}
\ee
and
\be
\lambda \tilde Z_1 F_\mu G^2_\mu = \frac{2\omega}{9} (\gamma+1) \,.
\mlab{uvcond2'}
\ee

Substituting the solutions Eqs.~(\oref{Fuvsol}, \oref{Guvsol}) in
the gluon equation \mref{Feqsub} and keeping only the leading log
terms, we now find
\ba
F^{-1}_\mu  \, \l[\omega\log\l(\frac{x}{\mu}\r)+1\r]^{-\gamma}
&=& F^{-1}_\mu  \, \l[\omega\log\l(\frac{\sigma}{\mu}\r)+1\r]^{-\gamma} \\
&& \hspace{-3cm} -7\lambda Z_1 F^2_\mu \,
\int_x^\sigma \, \frac{dy}{y} \,  \, 
\l[\omega\log\l(\frac{y}{\mu}\r)+1\r]^{2\gamma}
+ \lambda \tilde Z_1 G^2_\mu \,
\int_x^{\sigma} \, \frac{dy}{2y} \, 
\l[\omega\log\l(\frac{y}{\mu}\r)+1\r]^{2\delta} \,. \nn
\ea

After evaluation of the integrals and substitution of \mref{uvcond1},
\ba
F^{-1}_\mu  \, \l[\omega\log\l(\frac{x}{\mu}\r)+1\r]^{-\gamma}
&=& F^{-1}_\mu  \, \l[\omega\log\l(\frac{\sigma}{\mu}\r)+1\r]^{-\gamma} 
\mlab{FUV3} \\
&-& \frac{7\lambda Z_1 F^2_\mu}{\omega(2\gamma+1)}
\l\{\l[\omega\log\l(\frac{\sigma}{\mu}\r)+1\r]^{2\gamma+1}
- \l[\omega\log\l(\frac{x}{\mu}\r)+1\r]^{2\gamma+1}\r\} \nn\\
&-& \frac{\lambda \tilde Z_1 G^2_\mu }{2\omega\gamma} 
\l\{ \l[\omega\log\l(\frac{\sigma}{\mu}\r)+1\r]^{-\gamma}
- \l[\omega\log\l(\frac{x}{\mu}\r)+1\r]^{-\gamma} \r\} \,. \nn
\ea

Consistency of this equation requires $\gamma \le -1/3$, in order to equate
the leading log terms on both sides of the equation.  We first consider the
case $\gamma < -1/3$, for which the gluon loop does not contribute to
leading log. Then, the consistency of \mref{FUV3} requires that
\be
\lambda \tilde Z_1 F_\mu G^2_\mu = 2\omega\gamma \,.
\mlab{uvcond3}
\ee

From Eqs.~(\oref{uvcond2'}, \oref{uvcond3}) we then find
\be
\gamma = \frac{1}{8} \not< -\frac{1}{3} \,,
\ee
which is inconsistent with the initial assumption $\gamma < -1/3$.  The
only possibility left is
\be
\gamma = -1/3 \, ,
\ee
for which both the gluon and the ghost loop contribute to leading
order. From \mref{uvcond1} we then also find
\be
\delta=-1/3 \,,
\ee
and the condition \mref{uvcond2'} derived from the ghost equation yields
\be
\omega = \frac{27}{4} \lambda \tilde Z_1 F_\mu G^2_\mu \,.
\mlab{uvcond2''}
\ee

\mref{FUV3} then becomes
\ba
F^{-1}_\mu  \, \l[\omega\log\l(\frac{x}{\mu}\r)+1\r]^{1/3}
&=& F^{-1}_\mu  \, \l[\omega\log\l(\frac{\sigma}{\mu}\r)+1\r]^{1/3} 
\mlab{FUV4} \\
&& \hspace{-2.5cm} + \frac{\lambda}{\omega}\l[- 21 Z_1 F^2_\mu
+ \frac{3}{2} \tilde Z_1 G^2_\mu \r]
\l\{\l[\omega\log\l(\frac{\sigma}{\mu}\r)+1\r]^{1/3}
- \l[\omega\log\l(\frac{x}{\mu}\r)+1\r]^{1/3}\r\} \,, \nn
\ea
which leads to the condition:
\be
\omega = \lambda \l[21 Z_1 F^3_\mu
- \frac{3}{2} \tilde Z_1 F_\mu G^2_\mu \r] \,.
\mlab{uvcond4}
\ee

Eqs.~(\oref{uvcond2''}, \oref{uvcond4}) give us
\be
G^2_\mu = \frac{28 Z_1}{11 \tilde Z_1} F^2_\mu \,,
\mlab{uvcond5}
\ee
which is a relation between the leading-log renormalized values
of $F_\mu$ and $G_\mu$, when the renormalization scale $\mu$ is in
the perturbative regime, in which the leading log dominates. This
might seem to be in contradiction to perturbation theory, where the
values of the renormalized quantities can take an arbitrary value and
are usually fixed to 1. However, \mref{uvcond5} still contains the
renormalization constants $Z_1$ and $\tilde Z_1$. Taylor has
shown that $\tilde Z_1 \equiv 1$ in the Landau
gauge\cite{Taylor}, but one could still hope to be able to achieve
the arbitrary renormalization of $F$ and $G$ by a suitable choice of
$Z_1$.

If we write the far UV behaviour of $F(x)$ and $G(x)$ as
\be
F(x) \sim C \log^{-1/3}{x} \hspace{1cm}\mbox{and}\hspace{1cm}
G(x) \sim D \log^{-1/3}{x} \,,
\ee
then, from Eqs.~(\oref{Fuvsol}, \oref{Guvsol}, \oref{uvcond2''},
\oref{uvcond5}), the log-coefficients $C$ and $D$ of $F(x)$ and $G(x)$ are
given by
\be
C = F_\mu \, \omega^{-1/3} = \frac{1}{3} \l(
\frac{7 \lambda Z_1}{11} \r)^{-1/3}
\ee
and
\be
D = G_\mu \, \omega^{-1/3} = \frac{2}{3}\l(\frac{11 \lambda^2
\tilde Z_1^3}{7 Z_1}\r)^{-1/6}.
\ee

It is interesting to note that these leading log-coefficients are
independent of the values $F_\mu$ and $G_\mu$.

Let us now look at the ultraviolet behaviour of the running coupling. Using
the solutions Eqs.~(\oref{Fuvsol}, \oref{Guvsol})
and substituting \mref{uvcond2''}, we find
\be
\alpha F(x) G^2(x) = \frac{\alpha F_\mu G^2_\mu}
{\frac{27}{4} \lambda \tilde Z_1 F_\mu G^2_\mu 
\log\l(\frac{x}{\mu}\r)+1} \,.
\ee
Now, divide numerator and denominator by $\lambda F_\mu G^2_\mu$:
\be
\alpha F(x) G^2(x) = \frac{4\pi}
{\frac{27}{4} \tilde Z_1 \log\l(\frac{x}{\mu}\r)
+ \frac{1}{\lambda F_\mu G^2_\mu }} \,,
\ee
which can be written in the familiar form
\be
\alpha F(x) G^2(x) = \frac{4\pi}
{\beta_0 \log\l(\frac{x}{\Lambda^2_{QCD}}\r)} \,.
\ee

The leading-log coefficient $\beta_0=27/4$ if $\tilde Z_1 = 1$, which is
not in agreement with perturbation theory, for which $\beta_0=11$. This
seems somewhat puzzling, because all the perturbative ingredients are
contained in the non-perturbative equations, and the leading-log
perturbative result can be retrieved from a perturbative expansion of our
truncated set of Dyson-Schwinger equations. Nevertheless these perturbative
solutions are not consistent ultraviolet asymptotic solutions of the
non-perturbative equations themselves. The blame for this has to be put on
the specific truncation, which loses physical information about the gauge
theory. This is in contrast to the {\it ghost-loop-only} truncation, where
the non-perturbative $\beta$-coefficient is identical to the perturbative
one, when only including ghost loops in the full gluon propagator, and
which ultimately seems to have more physical relevance.

\subsection{Perturbative expansion}

Because of the previous remark, it is interesting to have a look at
the leading order perturbative expansion of the truncated set of
renormalized equations.  To perform the expansion we set the zeroth
order values as $F_0(x) = F_{\mu}$ and $G_0(x) = G_{\mu}$ and
substitute these constant values of $F$ and $G$ in the integrals of
\mref{Feqsub} and \mref{Geqsub}.  All the integrals can be solved
analytically and yield (for $\Lambda^2\to\infty$)
\be
F(x) = \frac{F_{\mu}}{1 + \lambda \l(7 Z_1 F^3_{\mu} -
\frac{1}{2}\tilde Z_1 F_{\mu}G^2_{\mu}\r)\log(x/\mu)} \,,
\ee
and analogously for the ghost form factor
\be G(x) = \frac{G(\mu)}{1 + \frac{9}{4}\lambda \tilde
Z_1 F_{\mu}G^2_{\mu}\log(x/\mu)} \,.
\ee

Combining these expressions and expanding the denominator to leading order
we obtain
\be \alpha F(x) G^2(x) = \frac{\alpha F_{\mu}G^2_{\mu}}{1
+ \lambda \l(7 Z_1 F^3_{\mu} + 4 \tilde Z_1 F_{\mu}G^2_{\mu}\r)
\log(x/\mu)} \,.
\mlab{pert1}
\ee

We note that taking $F_{\mu}=G_{\mu}$ and $Z_1=\tilde Z_1=1$
in the previous equation would
produce the perturbative result with $\beta_0=11$. However,
consistency with the non-perturbative integral equation does not allow such
a choice, because of \mref{uvcond5}. We can even check that, using the
condition \mref{uvcond5} in \mref{pert1}, we retrieve the coefficient
$\beta_0=27/4$, which means that using \mref{uvcond5} we find a
perturbative expansion in agreement with the non-perturbative equation,
although it is not in agreement with the usual result of pure perturbation
theory. This is an interesting observation which deserves future
research.
 
\subsection{Results}

We solved Eqs.~(\oref{Feqsub}, \oref{Geqsub}) with $\lambda=1$ and
$Z_1=\tilde Z_1=1$, for widely varying values of the parameters $A$ and
$F_1$ in order to scan the two-parameter space of solutions for a
given $\lambda$. However, the loss of symmetry seems to cut out part of
the solution space. Although we have made a rather thorough investigation
of this, we will not swamp the paper with a detailed discussion since we
think that this loss of symmetry is unphysical, thus making this truncation
less interesting than the {\it ghost-loop-only} truncation. To put it
briefly, the fact that the ultraviolet behaviours of $F(x)$ and $G(x)$ are
independent of $F(\mu)$ and $G(\mu)$ destroys the invariance of the running
coupling with respect to the choice of the individual renormalizations of
$F$ and $G$. Hence, choosing $F(\mu)$ too large will prohibit the
construction of a consistent solution having the correct ultraviolet
asymptotic behaviour. For $F(\mu)$ small this is not an obstacle, as we can
show that there is an intermediate regime, where the log of momentum takes
a different power, which allows us to connect to the correct ultraviolet
behaviour.

\begin{figure}
\begin{center}
\input{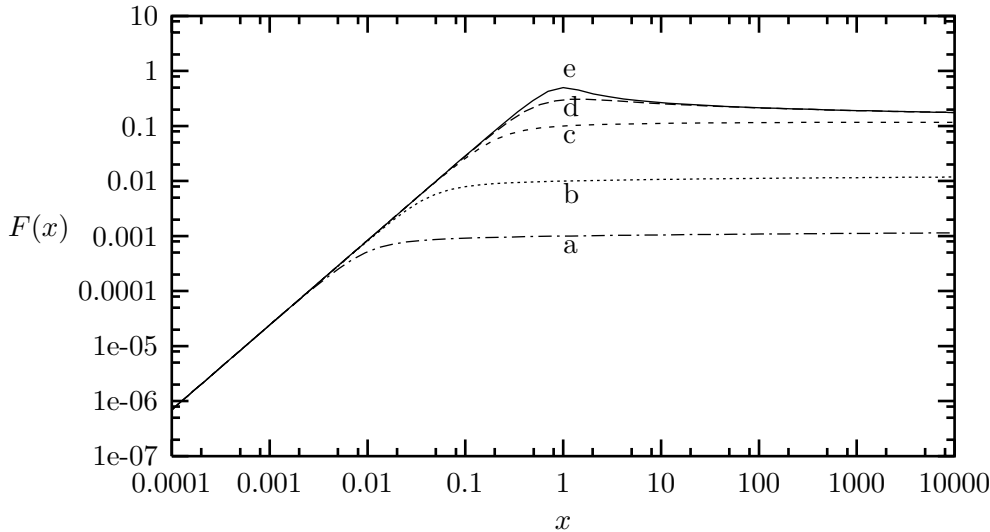}
\end{center}
\vspace{-1cm}
\caption{Gluon form factor $F(x)$ versus momentum $x$
(on log-log plot), for $A=1$ and $F_1=$ 0.001(a), 0.01(b), 0.1(c), 0.3(d)
and 0.5(e).}
\label{Fig:plot5}
\end{figure}

To see this we plot $F(x)$ for $A=1$ and $F_1=$ 0.001, 0.01, 0.1, 0.3 and
0.5 in Fig.~\ref{Fig:plot5}.
It is clear that the gluon form factor, which starts off as a power $A
x^{2\kappa}$, bends over at the cross-over point $\tilde x$, such that the
further logarithmic behaviour of the function leads to a value $F_1$ at the
renormalization scale $x=1$. From this plot it is however clear that
the curves (d, e) have a quite different behaviour from the others. 
Their ultraviolet behaviour is consistent with the $\log^{-1/3}$
analytic prediction from Sect.~\ref{Sect:UV}, while the other curves
seem to show a logarithmic increase instead. This is of course plausible,
as it is possible that the ultraviolet asymptotic behaviour only sets in at
much higher momenta, and that in between the infrared and ultraviolet
asymptotic behaviours there is a intermediate regime.

From a careful investigation of the equations, we can even find a
consistent analytical description of the intermediate regime, connecting
the region of confinement to that of asymptotic freedom, which fits the
numerical results extremely well. Consider a case where $|F(\sigma)| \ll
|G(\sigma)|$. Then, in the intermediate region, $F(x)G^2(x) \gg F^3(x)$, and
the gluon loop will be negligible compared to the ghost loop, in the gluon
equation, \mref{Feqsub}. Keeping in mind the treatment of
Sec.~\ref{Sect:UV-onlyghost}, we know that this has a consistent
ultraviolet solution $F(x) \sim \log^{1/8}{x}$ and $G(x) \sim
\log^{-9/16}{x}$, which remains valid all the way down to the region where
the power behaviour bends over to a logarithmic behaviour. Comparison with
the numerical results shows that indeed the intermediate regime is very
well reproduced by these powers of log. The ultimate ultraviolet behaviour
of Sect.~\ref{Sect:UV} will only set in at extremely high momentum, after
the intermediate regime has allowed the form factors to evolve
sufficiently in order to connect to the stringently constrained ultraviolet
asymptotic behaviour. The connection of the asymptotic infrared regime, the
intermediate log behaviour and the asymptotic ultraviolet behaviour
reproduce the numerical result to a good accuracy.

\begin{figure}
\begin{center}
\input{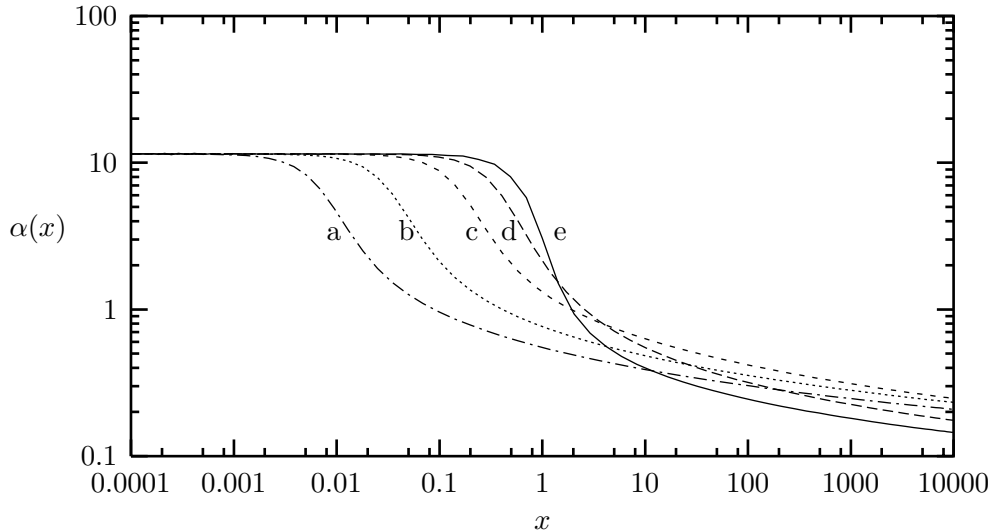}
\end{center}
\vspace{-1cm}
\caption{Running coupling $\alpha(x)$ versus momentum $x$
(on log-log plot), for $A=1$ and $F_1=$ 0.001(a), 0.01(b), 0.1(c), 0.3(d)
and 0.5(e).}
\label{Fig:plot4}
\end{figure}

To evaluate the physical relevance of this truncation, it is most
interesting to plot the running coupling in Fig.~\ref{Fig:plot4}.
We see that, in contrast to the {\it ghost-loop-only} truncation of
Fig.~\ref{Fig:plot1}, the various curves for the running coupling are no
longer mere translations of each other on log-log scale; and thus, if we
choose the units on each curve such that $\alpha(\mu) \equiv
\alpha_\mu^{\rm exp}$, we will find couplings which run in different ways
in the intermediate regime. This means that the determination of the
running of the strong coupling cannot be determined unambiguously in this
case.

\section{Conclusions}

Following the study of von Smekal et al.\cite{Smekal}, where these authors
studied the coupled system of Dyson-Schwinger equations for the gluon and
ghost propagators, using a Ball-Chiu vertex Ansatz for the triple gluon
vertex and a Slavnov-Taylor improved form for the gluon-gluon-ghost vertex,
we performed a detailed analytical and numerical analysis of the
coupled gluon-ghost equations using the {\it bare} triple gluon and
gluon-gluon-ghost vertices. The reasons to go back to the leading-order
perturbative vertices were to avoid the ad-hoc approximations von Smekal et
al. had to introduce in their integral equations in order to avoid
logarithmic infrared singularities and to obtain a clear understanding of
the mechanism that is the source of the new qualitative behaviour of the
non-perturbative gluon and ghost propagators and of the running
coupling. We believe that doing this has proven to be extremely
fruitful. Firstly, the qualitative changes to the infrared behaviours of
the propagators are solely due to the coupling of both propagator equations,
and the details of the vertices seem to introduce merely quantitative
changes. Secondly, the use of the bare vertices ensures that no infrared
singularities occur; hence no additional approximations, except for the
vertex Ans\"atze and the y-max approximation, are in principle
needed in order to solve these equations.

However, we did apply one more, physically motivated, truncation to the
coupled gluon-ghost equations. From an analysis of the symmetries of the
equations and their solutions, we believe that removing the gluon
loop, and keeping only the ghost loop in the gluon equation, leads to a set
of equations which is physically more relevant, because it is
consistent with the renormalization group invariance of the running
coupling, while this is not the case in the presence of the gluon loop (and
the approximations employed).

We performed a detailed analytical and numerical study of the equations
with and without the gluon loop. In the case where we removed the gluon
loop, we computed the analytical asymptotic infrared expansion, and showed
that it depends on three independent parameters defining the infrared
behaviour of a three-dimensional family of solutions. We also derived the
analytic ultraviolet asymptotic behaviour of the solutions, which are
proportional to powers of logarithms. We then computed the solutions for
$F(x)$, $G(x)$ and $\alpha(x)$ over the whole momentum range with two
different numerical techniques, the Runge-Kutta method on the set of
differential equations derived from the integral equations on the one hand,
and on the other hand, the direct solution of the integral equations using
a Newton iteration method to find Chebyshev approximations to the unknown
functions. The numerical results agree very well with both asymptotic
behaviours in the infrared and ultraviolet regions. Furthermore the results
of the Runge-Kutta method and of the direct integral equation method agree
to a very high accuracy. We found that the equations possess a
three-dimensional family of solutions and that they all correspond to one
and the same physical running coupling $\alpha(x)=\lambda F(x) G^2(x)$. The
non-perturbative running coupling can be matched unambiguously to physical
reality, and we showed how $\Lambda_{QCD}$ can be determined, at least in a
formal way, even though the $\beta$-coefficient is unphysical.

We repeated the study with inclusion of the gluon loop, and showed that
this truncation is physically less relevant and that the non-perturbative
running coupling cannot be determined unambiguously if one uses a bare
triple gluon vertex and takes $Z_1$ to be a constant.

To improve on the current study, we could try to reincorporate the gluon
loop in the gluon equation in a way that respects the physical invariances
of the problem. For this, we believe that the bare triple gluon vertex will
have to be replaced by an improved vertex, like the Ball-Chiu vertex, and
the renormalization constant $Z_1$ will have to be chosen
appropriately. Furthermore it would be interesting to investigate the
importance of the y-max approximation. 

{\bf Acknowledgements}

We thank A.~Hams for fruitful discussions. J.C.R.B. was supported by
F.O.M. ({\it Stichting voor Fundamenteel Onderzoek der Materie}).

\raggedright
\bibliographystyle{unsrt}
\bibliography{Biblio}

\begin{thebibliography}{10}

\bibitem{GtH}
G.~'t~Hooft, Nucl. Phys. {\bf B33} (1971) 173; {\bf B35} (1971) 167.

\bibitem{Politzer}
H.D.~Politzer, Phys. Rev. Lett. {\bf 30} (1973) 1346.

\bibitem{Roberts}
A useful overview can be found in C.D.~Roberts and A.G.~Williams, Prog. Part.
  Nucl. Phys. {\bf 33} (1994) 477, and references contained therein.

\bibitem{Mandelstam}
S.~Mandelstam, Phys. Rev. {\bf 20} (1979) 3223.

\bibitem{DA}
D.~Atkinson, J.K.~Drohm, P.W.~Johnson and K.~Stam, Jour. Math. Phys. {\bf 22}
  (1981) 2704.\\ D.~Atkinson, P.W.~Johnson and K.~Stam, Jour. Math. Phys. {\bf
  23} (1982) 1917.

\bibitem{MRP}
N.~Brown and M.R.~Pennington, Phys. Rev. {\bf D38} (1988) 2266; {\bf D39}
  (1989) 2723.\\ K.~B\"uttner and M.R.~Pennington, Phys. Rev. {\bf D52} (1995)
  5220.

\bibitem{Smekal}
L.~von~Smekal, A.~Hauck and R.~Alkofer, Phys. Rev. Lett. {\bf 79} (1997) 3591;
  hep-ph/9707327.

\bibitem{chiral}
D. Atkinson and P.W. Johnson, Phys. Rev. {\bf D37} (1988) 2290; {\bf D37}
  (1988) 2296.\\ D. Atkinson, P.W. Johnson and K. Stam, Phys. Rev. {\bf D37}
  (1988) 2996.

\bibitem{Taylor}
J.C.~Taylor, Nucl. Phys. {\bf B33} (1971) 436.

\bibitem{Bloch95b}
J.C.R.~Bloch, {\it Numerical Investigation of Fermion Mass Generation in QED},
  Ph.D. Thesis, University of Durham, 1995; J.C.R.~Bloch, in preparation.

\bibitem{PDG}
Particle Data Group, Phys. Rev. {\bf D54} (1996) 1.

\end{thebibliography}

\end{document}